\documentclass[universe,article,accept,pdftex,oneauthors]{Definitions/mdpi} 

\usepackage{epsfig} 

\def\be{\begin{equation}}
\def\ee{\end{equation}}
\def\ba{\begin{eqnarray}}
\def\ea{\end{eqnarray}}
\usepackage{color}

\def\b{\mathbf{b}}

\def\r{\bar{\rho}}
\firstpage{1} 
\pubvolume{1}
\issuenum{1}
\articlenumber{0}
\pubyear{2024}
\copyrightyear{2024}
\externaleditor{\textls[-25]{Academic Editor: Firstname Lastname}}
\datereceived{1 January 2024} 
\daterevised{30 January 2024} 
\dateaccepted{31 January 2024} 
\datepublished{ } 
\hreflink{https://doi.org/} 

\Title{Probing for Lorentz Invariance Violation in Pantheon Plus Dominated Cosmology}

\TitleCitation{Probing for Lorentz Invariance Violation in Pantheon Plus Dominated Cosmology}

\Author{{Denitsa Staicova} \orcidA{}}

\AuthorNames{Denitsa Staicova}

\AuthorCitation{{Staicova, D.}
}

\address[1]{Institute for Nuclear Research and Nuclear Energy, Bulgarian Academy of Sciences, {1784} Sofia, Bulgaria; dstaicova@inrne.bas.bg}

\abstract{The Hubble tension in cosmology is not showing signs of alleviation and thus, it is important to look for alternative approaches to it. One such example would be the eventual detection of a time delay between simultaneously emitted high-energy and low-energy photons in gamma-ray bursts (GRB). This would signal a possible Lorentz Invariance Violation (LIV) and in the case of non-zero quantum gravity time delay, it can be used to study cosmology as well. In this work, we use various astrophysical datasets (BAO, Pantheon Plus and the CMB distance priors), combined with two GRB time delay datasets with their respective models for the {\em {intrinsic time delay}}.  
Since the intrinsic time delay is considered the largest source of uncertainty in such studies, finding a better model is important. Our results yield as quantum gravity energy bound $E_{QG}\ge 10^{17}$ GeV and $E_{QG}\ge 10^{18}$ GeV respectively. The difference between standard approximation (constant intrinsic lag) and the extended (non-constant) approximations is minimal in most cases we conside. However, the biggest effect on the results comes from the prior on the parameter $\frac{c}{H_0 r_d}$, emphasizing once again that at current precision, {{cosmological datasets are} the dominant factor in determining the cosmology}. { We estimate the energies at which cosmology gets significantly affected by the time delay dataset}. 
}


\begin{document}
\section{Introduction}
\label{sec:intro}
The current cosmological probes have reached an unprecedented level of precision and understanding of the systematics related to measurements. Yet, the~unanswered questions remain, with~the tensions in cosmology the most famous among them. { {Currently,} the~Hubble tensions stands at $>5 \sigma$ \cite{Abdalla:2022yfr, Vagnozzi:2023nrq} and the need for new approaches is clear~\cite{Benisty:2022psx, Dainotti:2022bzg, Dias:2023qrl, Dialektopoulos:2023jam, Alonso:2023oro, Benisty:2023vbz, Dialektopoulos:2023dhb, Briffa:2023ern, Zhai:2023yny, Bernui:2023byc, Yang:2022kho, Gariazzo:2021qtg, Bargiacchi:2023jse, Staicova:2023vxe, Dainotti:2021pqg}}.

{ {The search for} Lorentz Invariance Violation (LIV) through astrophysical probes has a long history} \cite{Colladay:1996iz,  AmelinoCamelia:1997gz, Colladay:1998fq,Kostelecky:1988zi, Ellis:2005sjy, Jacob:2008bw, Gubitosi:2009eu, Vasileiou:2015wja, Amelino-Camelia:2013wha, Magueijo:2001cr, Amelino-Camelia:2000bxx, Amelino-Camelia:2000stu, Amelino-Camelia:2002cqb, Kostelecky:2008ts, Wei:2021ite, Wei:2021vvn, Zhou:2021ycu, Addazi:2021xuf, Abdalla:2022yfr, Desai:2023rkd,  Anchordoqui:2021gji, Abdalla:2023gmc,Pasumarti:2023ify,  Bolmont:2022yad, Rosati:2015pga, Amelino-Camelia:2020bvx, Pfeifer:2018pty, Amelino-Camelia:2022pja, Carmona:2022lyg, Bolmont:2014cma, MAGIC:2020egb, Lobo:2020qoa}. {Some quantum gravity theories predict violations of relativistic symmetries through messenger dispersion (photons, neutrinos, gravitational waves), the~detection of which might offer crucial clues for unified theories~\cite{Addazi:2021xuf}.}

\begin{figure}[H]
    \includegraphics[width=0.53\textwidth]{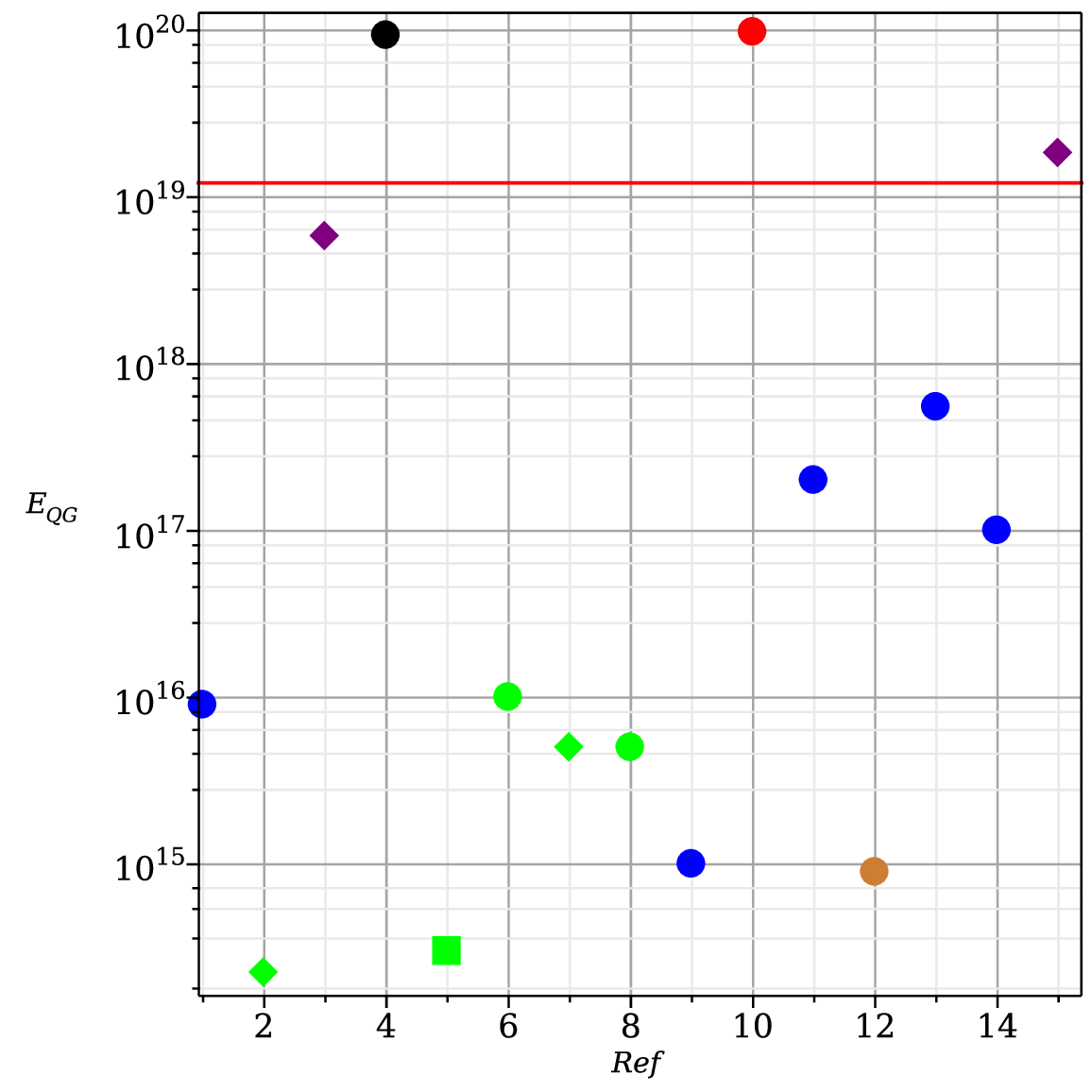}
    \caption{Current upper limits of $E_{QG}$:  P1~\cite{Ellis:2005sjy}, P2~\cite{Du:2020uev}, P3~\cite{MAGIC:2020egb}, P4~\cite{Vasileiou:2013vra}, P5~\cite{Pan:2020zbl}, P6~\cite{Agrawal:2021cim}, P7~\cite{Wei:2016exb}, P8~\cite{Desai:2022hht}, P9~\cite{Xiao:2022ovb}, P10~\cite{Chang:2012gq}, P11~\cite{Shao:2009bv}, P12~\cite{Vardanyan:2022ujc}, P13~\cite{Staicova:2023vln}, P14~\cite{Staicova:2023vln}, P15~\cite{LHAASO:2023lkv}. The~red line signifies the Planck energy. The legend can be found in the text.
    }
    \label{fig:LIV_Bounds}
\end{figure}

There are two possible ways to look for LIV---either locally by dedicated experiments~\cite{AlvesBatista:2023wqm}, that are so far out of our reach, or~alternatively, through cosmological probes. The~reason for this is that LIV effects are supposed to be amplified by the distance and also by the energy of the emission. Because~of this, astrophysical probes such as gamma-ray bursts (GRB) are very well suited for such studies. Gamma-ray bursts possess two important qualities for such studies---they can be seen at extreme distances ($z_{max}\sim 9.4$ \cite{Cucchiara:2011pj}) and at extreme energies ($E_{max} \sim 10^{55} erg$ \cite{Burns:2023oxn}). Additionally, GRB's emissions have been observed in a very wide energy band, spanning from keV to TeV (for example, GRB 221009A with emission >$10$ TeV~\cite{LHAASO:2023lkv,HESS:2023qhy}). LIV effects are usually measured by the bound of the quantum energy $E_{QG}$ above which they could be observed. A~plot of some measurements along with the references can be found in Figure~\ref{fig:LIV_Bounds}.  The~color legend of the figure refers to the different approximations for the intrinsic time delay used to obtain the points --- the blue square uses the standard approximation~\cite{Ellis:2005sjy, Shao:2009bv},  the~green color uses the energy fit~\cite{Wei:2016exb, Du:2020uev, Agrawal:2021cim, Desai:2022hht,Xiao:2022ovb}, the~red one uses the fireball model~\cite{Chang:2012gq}, the~brown one uses a variable luminosity~\cite{Vardanyan:2022ujc}, and~the black circle uses the SME framework~\cite{Vasileiou:2013vra}. The~most stringent bounds come from the TeV emissions {{of GRB 221009A} (18 TeV~\cite{LHAASO:2023lkv}) and GRB $190114C$ (0.2 TeV~\cite{MAGIC:2020egb})}.

While the eventual LIV effect would be very small, if~it is different from zero, it can also contribute to cosmology studies, providing new datasets independent of the luminosity measurements. Such datasets critically depend on the goodness of the GRB model (affecting the intrinsic time delay) and on the understanding of the propagational systematics of the messenger (affecting the other components of the time delay). {{Yet,} they could} help us look at cosmological tensions from another angle. { {The} question of the use of GRBs in cosmology has been studied extensively, for~example, using them as standardized candles~\cite{Dainotti:2022wli, Cao:2021irf, Dainotti:2022ked, Xu:2020moi} (and reference therein), through the cosmographic approach~\cite{Bargiacchi:2023rfd} and trying to reduce the Hubble tension~\cite{Dainotti:2023bwq}, in~combination with BAO, supernovae and quasars and a combined cosmological parameter~\cite{Staicova:2022zuh}}.

In a previous paper~\cite{Staicova:2023vln}, we started our investigation on how cosmology affects such constraints. LIV can be constrained either from the different {{energy}} bands of a single event or from averaging over multiple events, or~both. However,~cosmology needs to be taken into account in the estimations. It is particularly important when averaging over multiple GRBs due to the different redshifts employed. In~\cite{Staicova:2023vln}, we obtained that the effect of adding cosmology may be significant for certain models and datasets. The~biggest unknown in such a study is the intrinsic time delay---i.e.,~the possibility that the high and low-energy photons were not emitted at the same time. Since we do not have a good enough model of the GRB progenitor to predict the intrinsic time delay, we need to approximate it with a toy model. In~\cite{Staicova:2023vln}, we used the standard approximation that assumes a constant (over the energy or the luminosity) intrinsic time delay, common for all the~GRBs.

Here, we change that approximation, with~ two of the most popular other approximations. {{The} first one is the energy-dependent intrinsic time delay introduced in~\cite{Wei:2016exb, Ganguly:2017lwm}, which yields as bound for the quantum gravity energy scale  $E_{QG,1} \ge 0.3 \times 10^{15}$ GeV from GRB 160625B and 23 more GRBs~\cite{Pan:2020zbl} and $E_{QG}^1\ge 10^{16}$ GeV~\cite{Agrawal:2021cim}. The~other is the luminosity-dependent approximation for which the previously published result is $E_{QG}^1\ge 10^{15}$ GeV~\cite{Vardanyan:2022ujc}}.

In this paper, {{we} use these two extended approximations for the intrinsic time delay and we apply them to two available GRB time delay (TD) datasets, to~which we add several robust cosmological datasets. These are the angular} baryonic acoustic oscillations (transversal BAO), supernovae type IA (the Pantheon Plus dataset), and~the CMB distance prior.  To~the standard $\Lambda$CDM model, we add two dark energy (DE) models and {{a spatial} curvature model ($\Omega_K$CDM)}. We show that at the current level of precision for the LIV parameters, the~cosmology is more affected by the priors on the $c/H_0 r_d$ parameter than {{by}} the intrinsic time delay approximation. We see that indeed the extended approximation has an effect on our measurement of the LIV parameter $\alpha$, but in most cases it is small. The~other LIV parameters remain largely~unconstrained.

The paper is organized as follows: in Section~\ref{sec:theory}, we discuss the Theoretical background. In~Section~\ref{sec:methods}, we describe our Methods. In~Section~\ref{sec:datasets}, we elaborate on our Datasets. In \mbox{Section~\ref{sec:results}}, we discuss the numerical results, and in Section~\ref{sec:discussion}, one can find a Discussion of the obtained~results. 

\section{Theoretical~Setup}
\label{sec:theory}
\unskip
\subsection{Definition of the LIV Time~Delay}
We consider a modified dispersion relation for photons~\cite{AmelinoCamelia:1997gz, Vasileiou:2013vra} of the type: 
\be
E^2=p^2 c^2\left[1-s_\pm\left(
 \frac{E}{E_{QG,n}}\right)^n\,\right]\,,
 \ee
 with $E_{QG}$---{{the} energy of the QG scale},  $c$ is the speed of light, $p$ and $E$---the momentum and energy of photons and the $s_\pm=\pm 1$ for subluminal or superluminal propagation~\cite{Vasileiou:2013vra}. Here, we work only with $n=1$ and {{subluminal} LIV}.
 
Following~\cite{Jacob:2008bw,Biesiada:2007zzb, Biesiada:2009zz}, the~time delay between a low-energy ($E_{low}$) and a high-energy ($E_{high}$) photon with $\Delta E= E_{high}-E_{low}$ in the subluminal case is
\be
\Delta t_{LIV} = \frac{\Delta E}{E_{QG}}\int_0^z (1 + z' )\frac{dz'}{H(z')}
\ee

\noindent where $H(z) = H_0 E(z)$ is the Hubble parameter, $H_0$---the Hubble constant at $z=0$ and $E(z)$---the equation of state of the~Universe. 
\subsection{Different LIV~Approximations}

The observed time delay is a combination {{of}
} the quantum-gravity time delay, the~intrinsic time delay and different propagational effects~\cite{Zou:2017ksd, AlvesBatista:2021sln, Saveliev:2023urg}. While the propagational effects can be ignored for high-energy photons, to~put bounds on the quantum gravity time delay, we need to make an approximation for the intrinsic time delay. Here, we will compare the standard approximation with two of the most popular other~approximations.

 {1.}\textbf{{Standard approximation}} (e.g.,~\cite{Ellis:2005sjy,Biesiada:2009zz,Pan:2015cqa,Zou:2017ksd}). For~it, one assumes the following forms for the intrinsic time delay: $\Delta t_{\rm int}=\beta\left(1+z\right)$, where $\beta$ {{is}} a free~parameter.

 In this case, {{for} the GRB time delay, one obtains}:
 \be
 \frac{\Delta t_{obs}}{1+z}=a_{LIV}K+\beta\,,
 \label{eq:SA}
 \ee
 where $a_{LIV}\equiv\Delta E/(H_0 E_{QG})$, and~ \be
 K\equiv\frac{1}{1+z}
 \int_0^z\frac{(1+\tilde{z})\,d\tilde{z}}{E(\tilde{z})}\,.
 \ee
 If $a_{LIV}=0$, there is no LIV, while if $a_{LIV}\not=0$, there is LIV on energy scales above $E_{QG}$. {{The} connection between the QG effect and cosmology comes from the term $E(z)$}.

In our models, we use the form 
 \be
 \Delta t_{obs} = a_{LIV}(1+z)K + \beta\left(1+z\right)\,.
 \label{eq:tobs}
 \ee
{{As}
in our previous paper~\cite{Staicova:2023vln}, we denote $a_{LIV}=\alpha$ and $b$ is replaced with $\beta$, to~avoid mixing this parameter with $b=c/(H_0 r_d)$}.

{2.} \textbf{{The energy-dependent approximation}} This approximation has been used \linebreak\mbox{in~\cite{Wei:2016exb, Pan:2020zbl, Agrawal:2021cim, Desai:2022hht}} and also in~\cite{Du:2020uev} with ($\alpha\to-\alpha$). This popular formula assumes that in {the} source frame, the intrinsic positive time lag increases with the energy $E$ as a power law:

\be
\delta t_{int} = \tau \left [\left(\frac{E}{keV}\right)^\alpha- \left(\frac{E_0}{keV}\right)^\alpha \right]
\label{eq:EA1}
\ee

Here, $E_0=11.34$ keV is the median value of the fixed lowest energy band (10–12 keV). The~free parameters are $\tau$ and $\alpha$. While this fit is usually associated with studying a single GRB in multiple energy bands,~\cite{Desai:2022hht} has also applied it to multiple GRBs. Note, from~here on, we rename the parameters as follows: $\alpha \to \gamma, \tau \to \beta$. This is in order to differentiate them from the LIV $\alpha$ and the cosmological $b$.

{3.} \textbf{{Intrinsic GRB lag-luminosity relation approximation}}

This formula has been used in~\cite{Vardanyan:2022ujc} to describe the intrinsic time delay of long GRBs.
\be
\tau_{RF}^{int}=\tau_{obs}^{int}/(1+z)=\beta_{long}\left(\frac{L_{iso}}{L_*}\right)^{\gamma}
\label{eq:EA2}
\ee

Here, $L_*$ is arbitrary normalization, and~$\gamma$ and $\beta_{long}$ are the free parameters. For~short GRBs, one takes $\gamma=0$, which recovers the standard~approximation. 

\subsection{Cosmology}
{{The} equation of state of the Universe for a flat Friedmann-Lema\^itre-Robertson-Walker metric with the scale parameter $a = 1/(1+z)$  is:}
\begin{equation}
    E(z)^2 =  \Omega_{m} (1+z)^3 + \Omega_{DE}(z) + \Omega_K (1+z)^2,
    \label{eq:hz}
\end{equation}

\noindent where $\Omega_{DE}(z)\to \Omega_\Lambda$ for $\Lambda$CDM. The~ expansion of the universe is {{governed}} by $E(z)= H(z)/H_0$, where $H(z) := \dot{a}/a$ is the Hubble parameter at redshift $z$. $\Omega_{m}$, $\Omega_K$ and $\Omega_{DE}$ are the fractional matter, spatial and dark energy densities at redshift $z=0$.

We will consider a few dynamical dark energy models: number of different DE models: Chevallier-Polarski-Linder (CPL,~\cite{Chevallier:2000qy,Linder:2005ne,Barger:2005sb}) and  Barboza-Alcaniz (BA~\cite{Barboza:2008rh, Escamilla-Rivera:2021boq}). {{The} relevant equations for $\Omega_{DE}(z)$ for Equation~(\ref{eq:hz}) can be found in Table~\ref{table:DE_models}. For~$w_0=-1, w_a=0$ one recovers $\Lambda$CDM}.

\begin{table}[H]
\caption{{{The DE models} we use in this work. The~references can be found in the text.}}

\label{table:DE_models}
\newcolumntype{C}{>{\centering\arraybackslash}X}
\setlength{\tabcolsep}{10mm}
\begin{tabularx}{\textwidth}{ccc}
			\toprule
			  \textbf{Model} & {\boldmath {$\Omega_{DE}(z)$}}
&  {\boldmath $w(z)$}  \\
   \midrule
             CPL  &  $  \Omega_{\Lambda} \times\exp\left[\int_0^{z} \frac{3(1+w(z')) dz'}{1+z'}\right] $ & $w_0 + w_a \frac{z}{z+1}$
             \\
             BA & $\Omega_{\Lambda} \times(1+z)^{3(1+w_0)}{(1+z^2)}^{\frac{3w_1}{2}}$ & $w_0+z\frac{1+z}{1+z^2}w_1$   \\
 
   \bottomrule
		\end{tabularx}
	
\end{table}

\textls[-15]{To infer the cosmological parameters, we need to define the angular diameter \mbox{distance $D_\textrm{A}$:}}


\be
D_\textrm{A}
=\frac{c}{(1+z) H_0 \sqrt{|\Omega_{K}|}  } \textrm{sinn}\left[|\Omega_{K}|^{1/2}\int_0^z \frac {dz'} {E(z')}\right]\ ,
\label{eq:DA}
\ee
where $\textrm{sinn}(x) \equiv \textrm{sin}(x)$, $x$, $\textrm{sinh}(x)$ for $\Omega_{K}<0$, $\Omega_{K}=0$, $\Omega_{K}>0$, respectively.

It connects to the transversal BAO measurements through the angular scale measurement $\theta_{BAO}(z)$:
\begin{equation}
\theta_{BAO}\left(z\right) = \frac{r_d}{\left(1+z\right)D_A(z)}.
\end{equation}

The Pantheon Plus datasets measure the distance modulus $\mu(z)$, which is related to the luminosity distance ($d_L = D_A(1+z)^2$) in Mpc through
\begin{equation}
        \mu_B (z) - M_B = 5 \log_{10} \left[ d_L(z)\right] + 25  \,,
\label{eq:dist_mod_def}
\end{equation}
where $M_B$ is the absolute~magnitude.

Finally, we include the CMB through the distance priors datapoints provided by~\cite{Chen:2018dbv}:
\begin{align*}
 l_\textrm{A} =(1+z_*)\frac{\pi D_\textrm{A}(z_*)}{r_s(z_*)} ,\\
R\equiv(1+z_*)\frac{D_\textrm{A}(z_*) \sqrt{\Omega_m } H_0}{c},
\label{la:Rz}
\end{align*}
where $l_\textrm{A}$ is the acoustic scale {{coming} from} the CMB temperature power spectrum in the transverse direction and $R$ is the ``shift parameter'' {{obtained} from} the CMB temperature spectrum along the line-of-sight direction~\cite{Komatsu:2008hk}. $r_s(z_*)$ is the co-moving sound horizon at redshift photon decoupling $z_*$ ( $z_* \approx 1089$ \cite{Aghanim:2018eyx}).

\section{Methods}
\label{sec:methods}
The idea of the method is to avoid setting priors on $H_0$ and $r_d$ by~considering only the quantity $b=c/(H_0 r_d)$. The~method has been outlined in~\cite{Staicova:2023vln}. For this reason, we just list the different likelihoods here. For~BAO we have
\begin{equation}
\chi^2_{BAO} = \sum_{i} \frac{\left(\vec{v}_{obs} - \vec{v}_{model}\right)^2}{\sigma^2}.
\label{eq:chi_bao}
\end{equation}

{Here,} $\vec{v}_{obs}$ is a vector of the observed points, {{for} BAO corresponding to $\theta_{BAO, obs}^i$,   $\vec{v}_{model}$ is the theoretical prediction of the model (for BAO ---$\theta_{BAO, theor}(z_i)$)} and $\sigma$ is the error of each~measurement.

For the SN dataset, we marginalize over $H_0$ and $M_B$. The~integrated $\chi^2$ in this case \mbox{is (\cite{DiPietro:2002cz,Nesseris:2004wj,Perivolaropoulos:2004yr,Lazkoz:2005sp}}{):}
\begin{equation}
\tilde{\chi}^2_{SN,  GRB} = D-\frac{E^2}{F} + \ln\frac{F}{2\pi},
\end{equation}
for
\begin{subequations}
\begin{equation*}
D = \sum_i \left( \Delta\mu \, C^{-1}_{cov} \, \Delta\mu^T \right)^2,
\end{equation*}
\begin{equation*}
E = \sum_i \left( \Delta\mu \, C^{-1}_{cov} \, \boldsymbol{\mathit{E}} \right),
\end{equation*}
\begin{equation*}
F = \sum_i  C^{-1}_{cov}  ,
\end{equation*}
\end{subequations}
where  $\mu_{}^{i}$ is the observational {{distance} modulus}, $\sigma_i$ is its error, and~the $d_L(z)$ is the luminosity distance, $\Delta\mu =\mu_{}^{i} - 5 \log_{10}\left[d_L(z_i)\right)$, $\boldsymbol{\mathit{E}}$ is the unit matrix, and~$C^{-1}_{cov}$ is the inverse covariance matrix of the Pantheon Plus dataset as given by~\cite{Scolnic:2021amr, Brout:2022vxf}.

Finally, we define the time delay likelihood as Equation~(\ref{eq:chi_bao}), but here, the quantity we consider is the theoretical time delay ($\vec{\Delta t}_{model}$), as defined in Equation~(\ref{eq:tobs}) and its observational value (($\vec{\Delta t}_{obs}$), which is provided by the TD dataset. 
\begin{equation}
\chi^2_{TD} = \sum_{i} \frac{\left(\vec{\Delta t}_{obs} - \vec{\Delta t}_{model}\right)^2}{\sigma^2},
\label{eq:chi_TD}
\end{equation}

The final $\chi^2$ is
$$\chi^2=\chi^2_{BAO}+\chi^2_{CMB}+\chi^2_{SN}+\chi^2_{TD}$$

\section{Datasets}
\label{sec:datasets}
In this work, we use the so-called transversal BAO dataset published by~\cite{Nunes:2020hzy}, {{for} which the authors claim to be cleaned up from the dependence on the cosmological model and uncorrelated}. The~CMB distant prior is given by~\cite{Chen:2018dbv}. The~SN data comes from the  Pantheon Plus dataset. It consists of 1701 light curves of 1550 spectroscopically confirmed Type Ia supernovae and their covariances, from which distance modulus measurements have been obtained~\cite{Riess:2021jrx, Brout:2022vxf, Scolnic:2021amr}.

To study the time delays, we use two different time delays (TD) datasets---TD1 provided by~\cite{Vardanyan:2022ujc} and TD2~\cite{Xiao:2022ovb}. TD1 uses a combined sample of 49 long and short GRBs observed by Swift, dating between 2005 to 2013. In~this dataset~\cite{Bernardini:2014oya, Vardanyan:2022ujc}, the~time lags have been extracted through a discrete cross-correlation function (CCF) analysis  between characteristic rest-frame energy bands of 100–150 keV and 200–250 keV. The~redshift for TD1 is $z\in [0.35, 5.47]$.  TD2~\cite{Xiao:2022ovb} uses 46 short GRBs with measured redshifts at Fixed Energy Bands (15–70 keV and 120–250 keV) gathered between 2004 and 2020 by Swift/BAT or Fermi/GBM. The~two datasets have only six common GRBs, which is under 15\% of their total number (49 vs. 46 events), which makes them effectively uncorrelated and independent. Because~we want to emphasize the effect of cosmology, we prefer to average over multiple GRBs rather than to use measurements from a single GRB ~\cite{Ellis:2005sjy}.

To run the inference, we use a nested sampler, {{provided} by the} open-source package {$Polychord$}
 \cite{Handley:2015fda} and the {{package}} $GetDist$ package~\cite{Lewis:2019xzd} {{for} the plots}.

{{We} use uniform priors for all quantities}: $\Omega_{m} \in [0.1, 1.]$, $\Omega_{\Lambda}\in[0., 1 - \Omega_{m}]$, $c/ (H_0 r_d) \in [25, 35]$, $w_0 \in [-1.5, -0.5]$ and $w_a \in [-0.5, 0.5]$. Since the distance prior is defined at the decoupling epoch ($z_*$) and the BAO---at drag epoch ($z_d$), we parametrize the difference between $r_s(z_*)$ and $r_s(z_d)$ as $rat= r_*/r_d$, where the prior for the ratio is $rat \in [0.9, 1.1]$ . {{The}  LIV priors are $\alpha \in [0, 0.1], \beta \in [-1,1], \gamma \in [-3,0]$.}

\section{Results}
\label{sec:results}
In~\cite{Staicova:2023vln}, we started investigating the effect of cosmology on LIV bounds by studying in the standard approximation (Equation (\ref{eq:SA})) two databases: the most famous and robust time delay dataset published by Ellis~et~al.~\cite{Ellis:2005sjy} and the one that we currently refer to as TD1. To~avoid repetition. In~this section, we present only the extended approximations\endnote{When we refer to the ``extended model'', we will mean the Intrinsic GRB lag-luminosity relation (Equation \eqref{eq:EA1}) approximation for TD1 and the Energy-dependent approximation (Equation \eqref{eq:EA2}) for TD2.}. For~completeness, the~results for the standard approximation for TD1 and TD2 are presented in~{Appendix} \ref{apendix:3}.

Since our previous work~\cite{Staicova:2023vln} demonstrated a deviation from the expected values for $b$ and $\Omega_m$, here, we want to investigate further this observation. To~this end, we try two hypotheses for the two datasets---a uniform prior on $b$ and a Gaussian prior focused on the expected value from Planck results, $b=30 \pm 1$. For~the standard approximation, this leads to a very clear higher value of $\Omega_m$ but a minimal difference for the DE parameters $w_0$ and $w_a$ and $\Omega_K$. 

The results for the extended models are summarized~below. 
\begin{itemize}[leftmargin=8mm,labelsep=0.5mm]
\item[--] {\bf {Matter density}}
\end{itemize}

\begin{figure}[H]
    \includegraphics[width=0.43\textwidth]{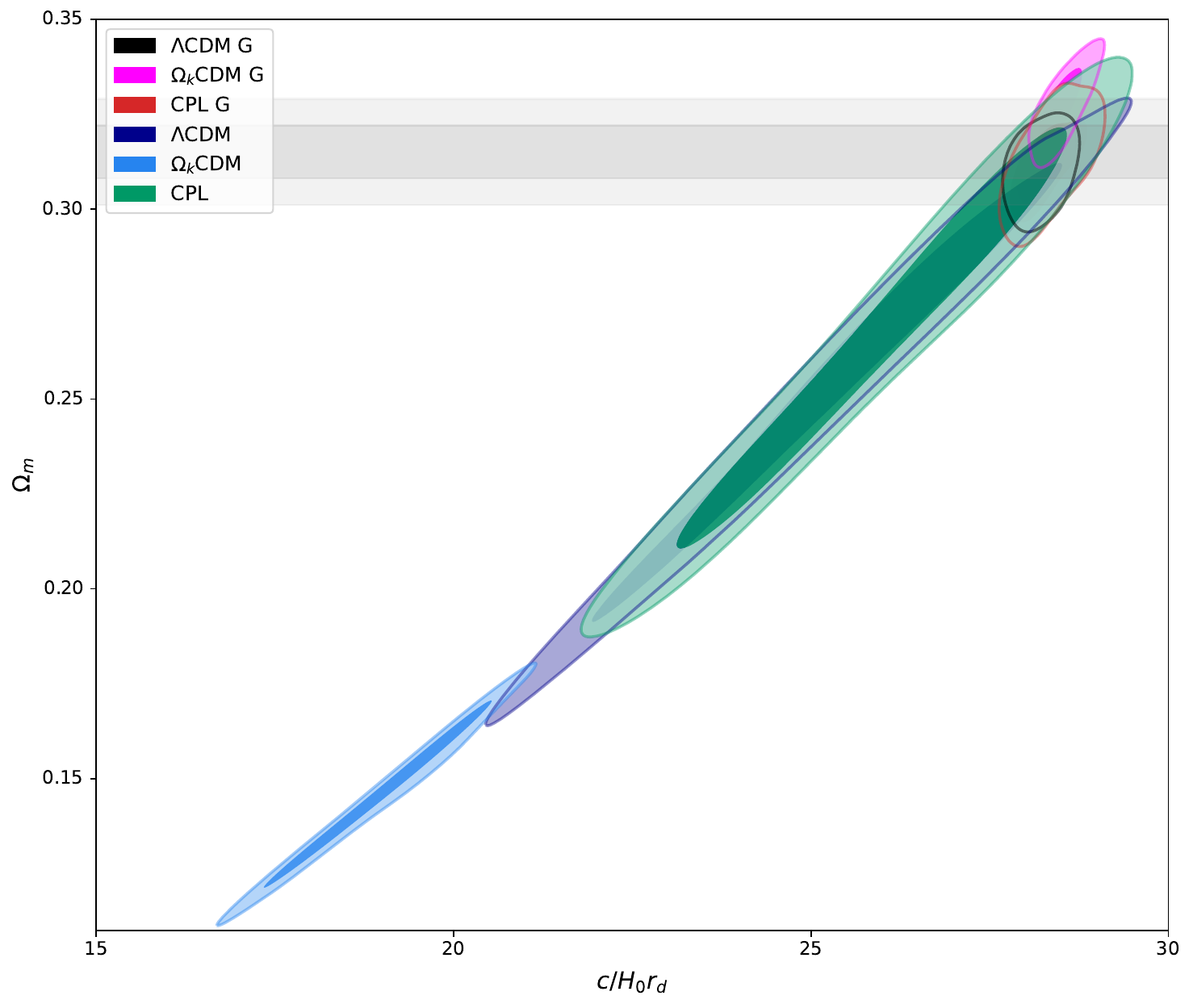}
    \includegraphics[width=0.43\textwidth]{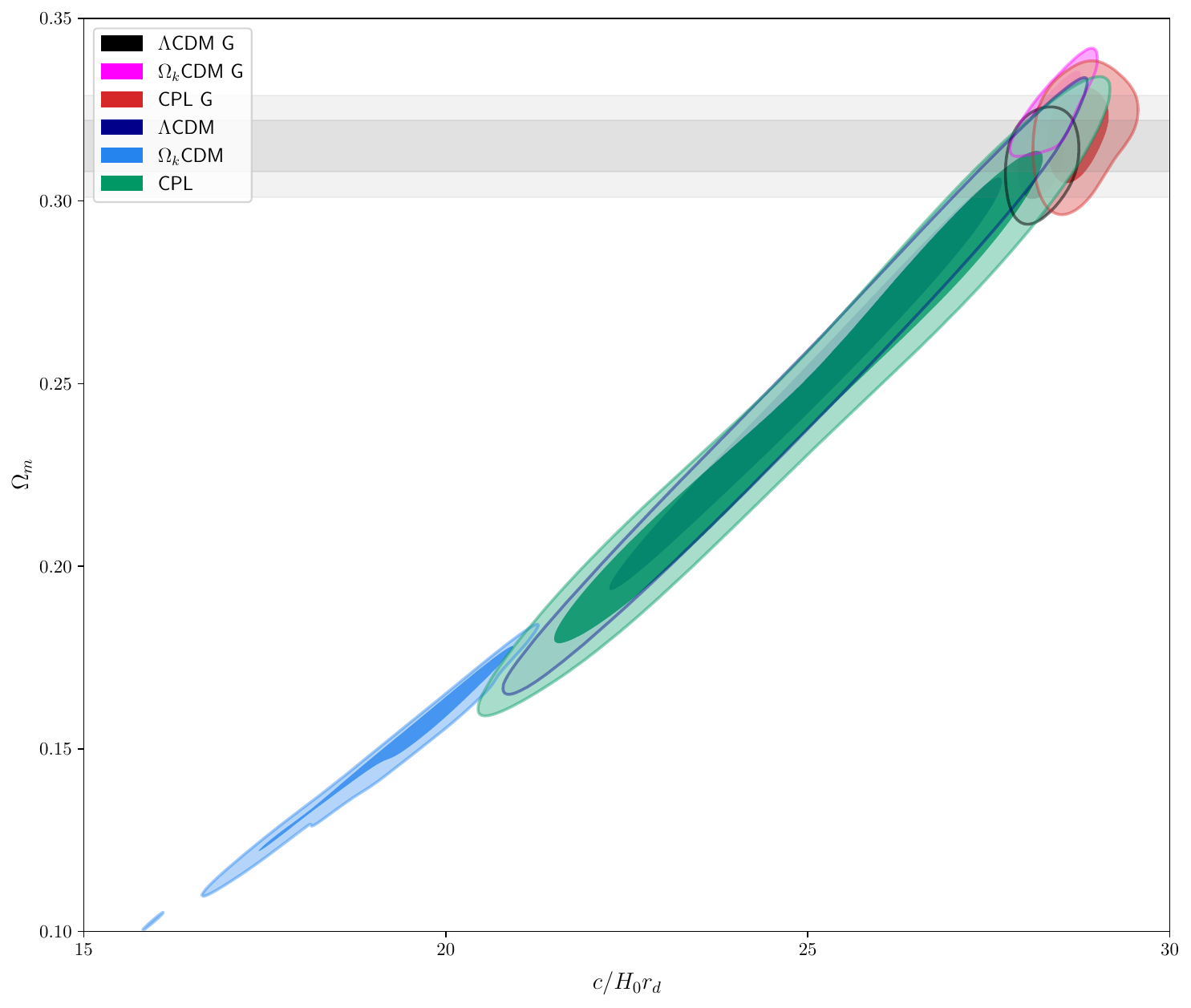}
    \caption{Posteriors {{(95}\%CL and 68\% CL)} for the TD1 dataset (\textbf{left}) and the TD2 dataset (\textbf{right}) for the parameters $b$ and $\Omega_m$. }
    \label{fig:bOm_L}
\end{figure}

In Figure~\ref{fig:bOm_L}, one can see the plot of $b$ vs. $\Omega_m$ for the extended LIV models for the \mbox{four different} cosmological models we consider (and the two priors on $b$).

For both datasets, we see two groups of posteriors---one grouped around the top right corner, which corresponds to the Gaussian priors on $b$, and~one that spans across the whole interval for the uniform prior.  {In both cases, the~mean value and the 95\% CL of $b$ are below the expected value of $b \sim 30$ for the three models.}

For TD1, setting the Gaussian prior for $b$ in this case significantly constrains $\Omega_m$ and puts it in the limits of [0.29,0.34] at 95\% CL. In~this case, $\Omega_k$CDM gives the highest mean $\Omega_m$ and $CPL$ gives the lowest one, with~$\Lambda$CDM in between. For~the uniform prior, the~posterior for $\Omega_m$ is much less constrained, with~the lowest value coming from $\Omega_K$CDM and the posteriors for $\Lambda$CDM and CPL largely~coinciding. 

For the other dataset, TD2, the~situation repeats to a great extent, with~a little bit higher matter density $\Omega_m \in [0.3,0.34]$ for the Gaussian prior and $b\in[ 27.8,29,2]$ and the lowest value for $\Omega_m$ is for $\Lambda$CDM and the highest for $\Omega_K$CDM. For~the uniform prior, we have again the lowest value for $\Omega_m$ and $b$ coming from $\Omega_K$CDM and $\Lambda$CDM and CPL largely~coinciding.

\begin{itemize}[leftmargin=8mm,labelsep=0.5mm]
\item[--] {\bf {Spatial curvature}}
\end{itemize}

\begin{figure}[H]
    \includegraphics[width=0.43\textwidth]{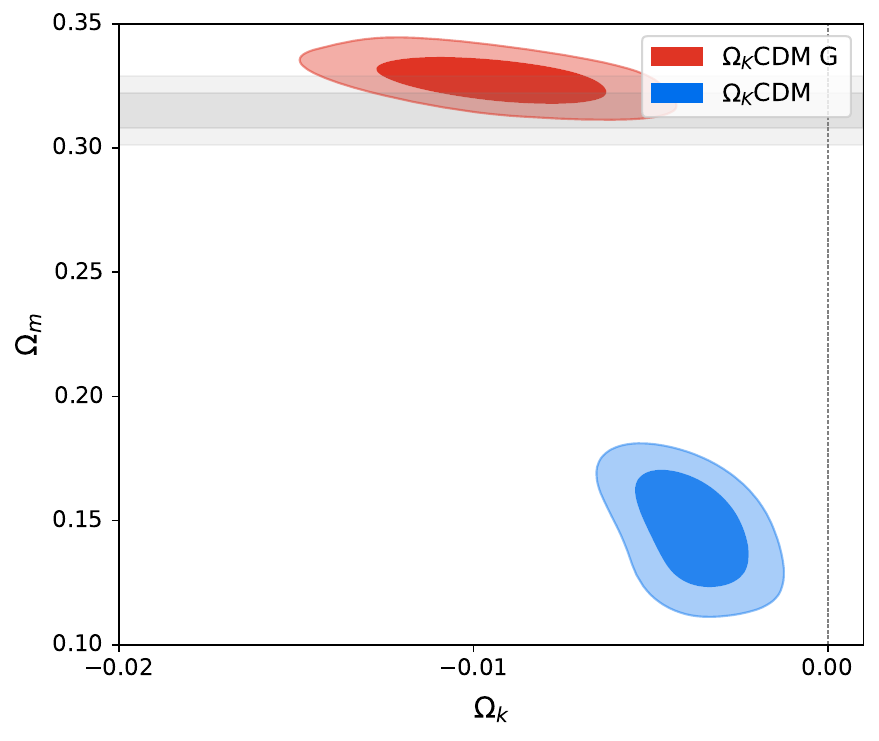}
    \includegraphics[width=0.43\textwidth]{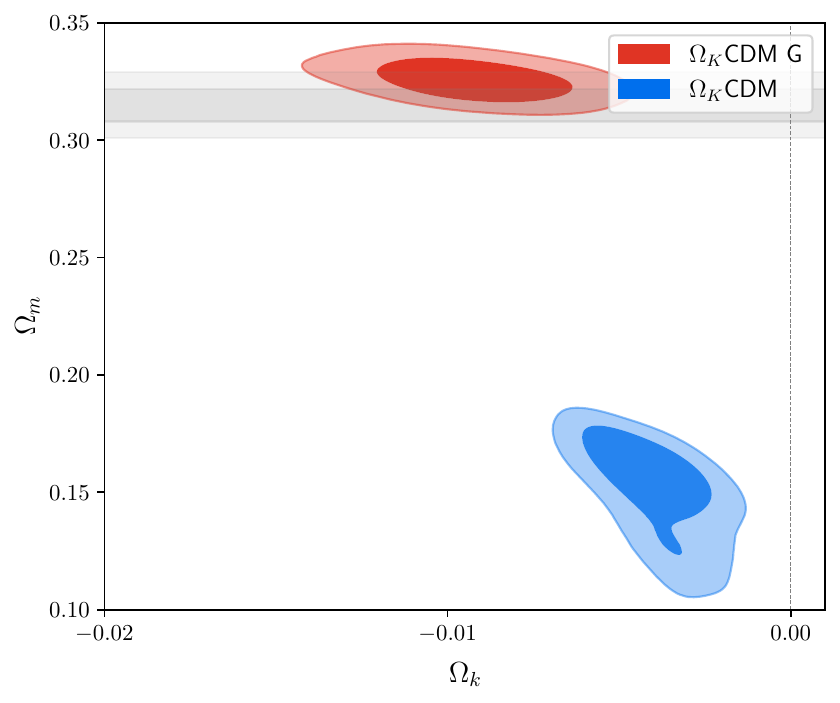}
    \caption{Posteriors  {{(95}\%CL and 68\% CL)} for the TD1 dataset (\textbf{left}) and the TD2 dataset (\textbf{right})---the spatial curvature $\Omega_K$ vs. $\Omega_m$.  }
    \label{fig:Vard_Xiao_Ok_L}
\end{figure}

The results for $\Omega_K$CDM are presented in Figure~\ref{fig:Vard_Xiao_Ok_L}.

We see that our results are not consistent with a flat universe at the 95\% CL even for a uniform prior on $b$. For~TD1, we get a bit less constrained contours for $\Omega_K$, but~more constrained than the flat cases. A~rather interesting feature is the huge difference in $\Omega_m$ between the uniform and the Gaussian cases for both datasets. { {Our} prior on $\Omega_K$ is relatively small ($\Omega_K\in[-0.1,0.1]$), however, due to the large number of parameters of the model. }

\newpage

\begin{itemize}[leftmargin=8mm,labelsep=0.5mm]
\item[--] {\bf {DE parameters}}
\end{itemize}

\begin{figure}[H]
    \includegraphics[width=0.43\textwidth]{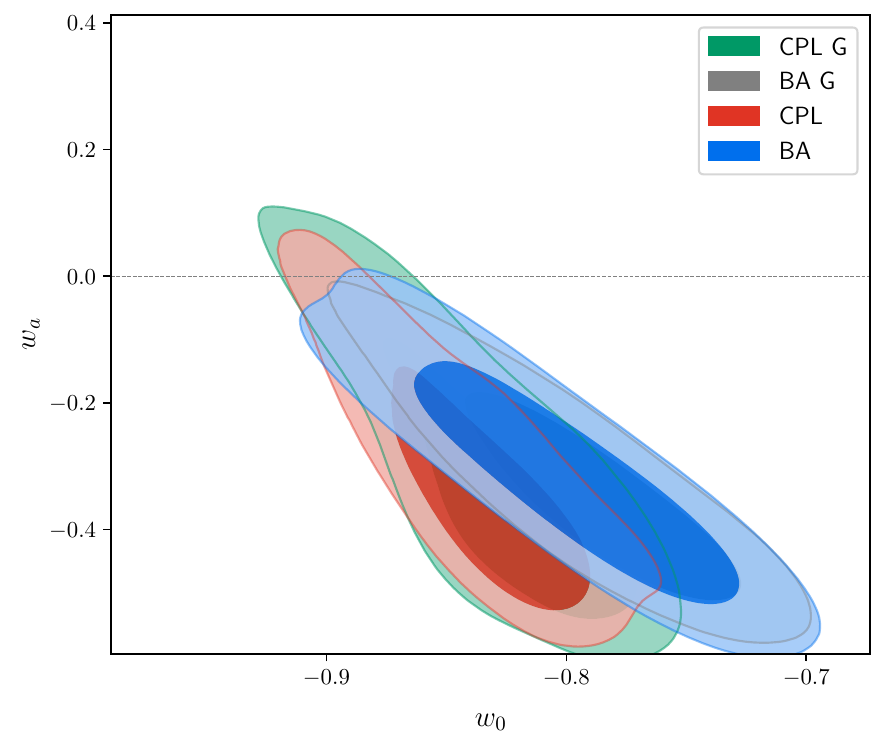}
    \includegraphics[width=0.43\textwidth]{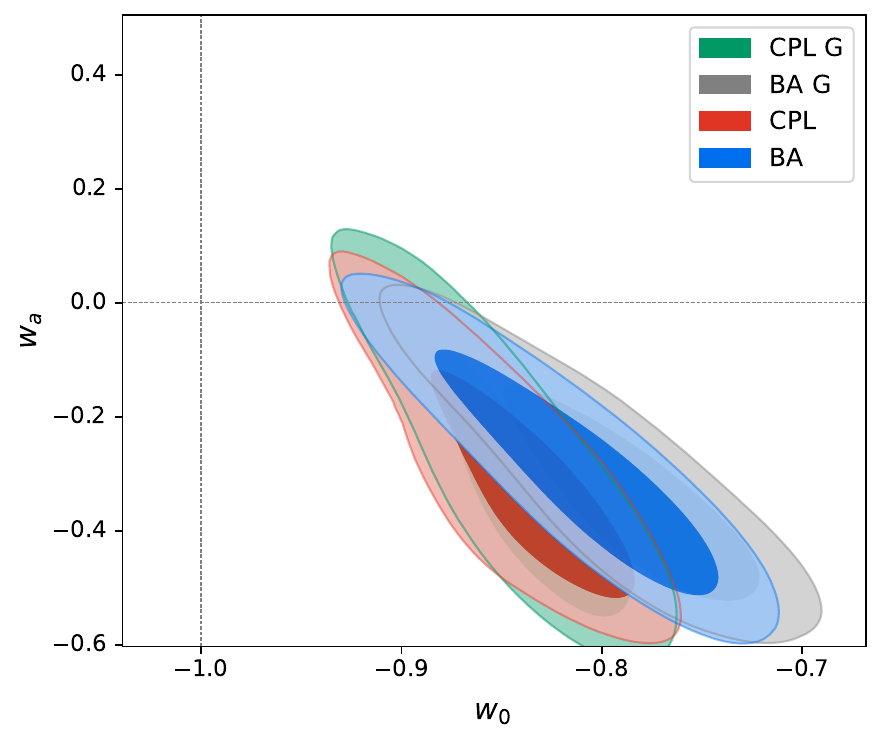}
    \caption{Posteriors {{(95}\%CL and 68\% CL)} for the TD1 dataset (\textbf{left}) and the TD2 dataset (\textbf{right}) in the $w_0-w_a$ plane.}
    \label{fig:Vard_Xiao_DE_L}
\end{figure}

The two DE models we consider are shown in Figure~\ref{fig:Vard_Xiao_DE_L}. Their DE parameters $w_0$ and $w_a$ are  largely insensitive to the dataset  and their mean values for $w_0$ are lower than expected for  $\Lambda$CDM, with~$w_0\sim -0.8$. This is a different value from the one obtained~\cite{Staicova:2023vln}, which was consistent with $\Lambda$CDM at 68\% CL. The~result in the study~\cite{Staicova:2022zuh} without TD on the other hand, shows  values about $w_0 \sim -1.1$. In~the two latter works, we use the Pantheon dataset, while here, we use the Pantheon Plus. To~clarify if the difference is due to the new SN dataset or to the new TD datasets, we ran the same experiment with Pantheon instead of Pantheon Plus and it gave $w_0\to -1$ within 68\% CL.  As~expected, the~parameter $w_0$ is well constrained by the data  while $w_a$ is not constrained at~all.

Note, the~cosmological results for TD1 and TD2 for the extended models look largely the same. That could be due to similar treatment of the data or other reasons (like adding additional degrees of freedom in the extended models). The~standard approximation posteriors presented in {Appendix} \ref{apendix:1} show a larger dependence on the dataset. Also, the~LIV contribution to the fit is quite small due to the smallness of $\alpha$. As~long as it is non-negligible, it still contributes to the cosmology in a minor way, { {as} seen above in $b$ and $\Omega_m$}. The~numerical experiment shows that for the TD1 dataset, the~effect of the TD dataset on the cosmology becomes really pronounced at about $\alpha\sim 10^{-2}$, while for TD2, it becomes pronounced at about $\alpha\sim 10^{-3}$. This is just an order or two above our current constraints for $\alpha$. This means that observing an event that would push the bound for $E_{QG}$ lower (or having a GRB intrinsic time delay model that would do so) would also bring the TD measurement on par with the current cosmological~probes.

\begin{itemize}[leftmargin=8mm,labelsep=0.5mm]
\item[--] {\bf {LIV Parameters}}
\end{itemize}

\begin{figure}[H]
    \includegraphics[width=0.47\textwidth]{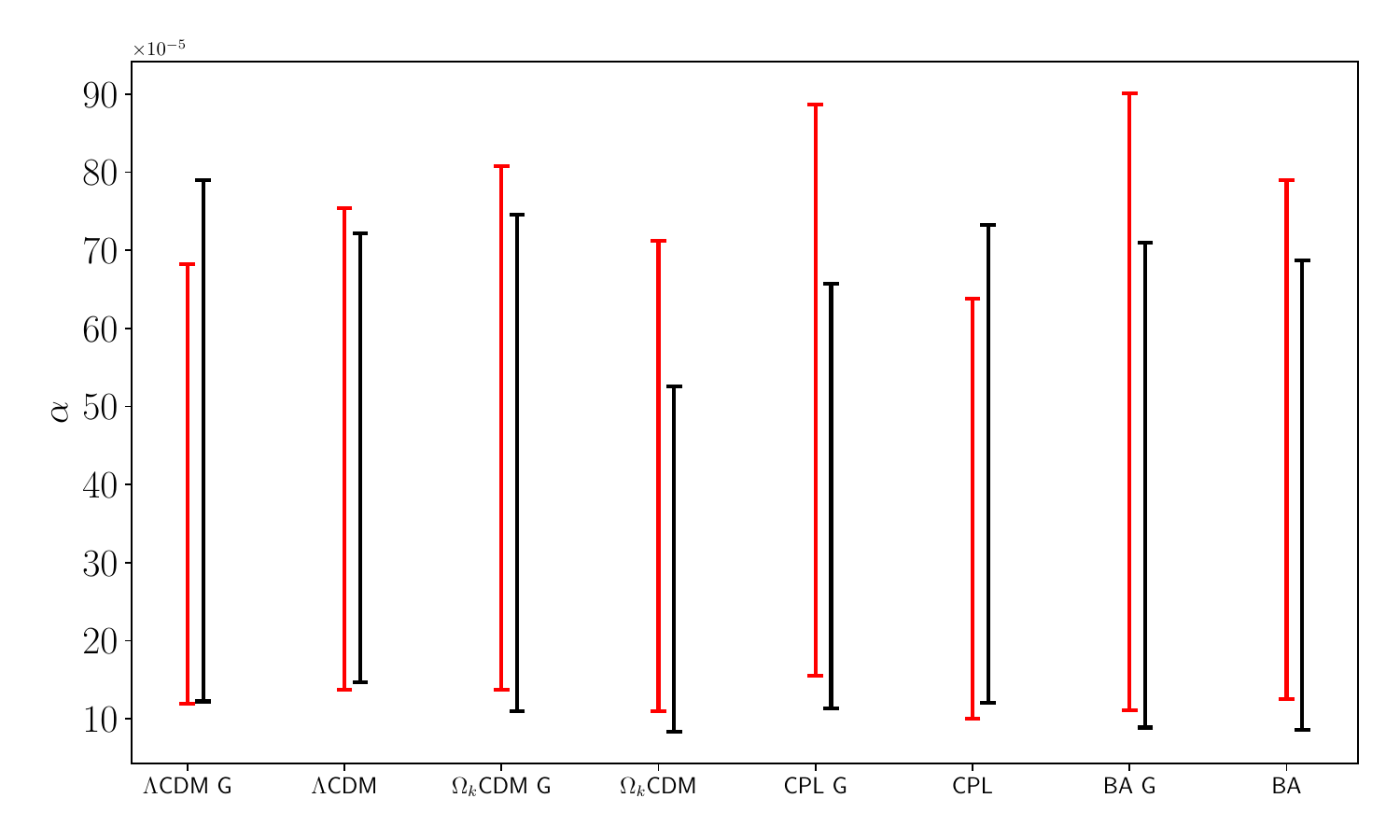}
    \includegraphics[width=0.47\textwidth]{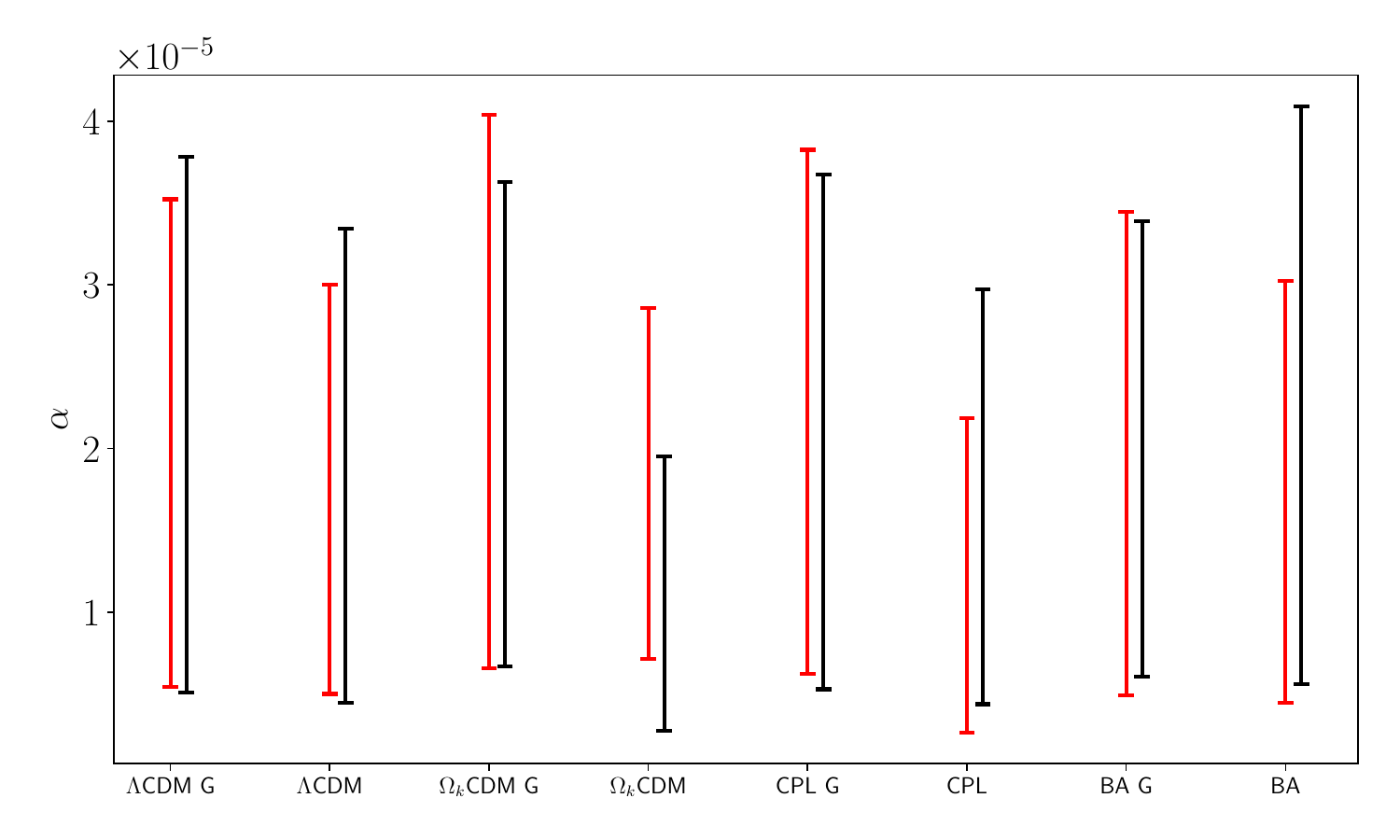}
    \caption{Mean values and errors for the parameter $\alpha$ for the standard approximation (red) and the extended one (black) different models. To~the (\textbf{left}) are the results for TD1 and to the (\textbf{right})---from~TD2.}
    \label{fig:Vard_Xiao_alpha}
\end{figure}

\begin{figure}[H]
    \includegraphics[width=0.47\textwidth]{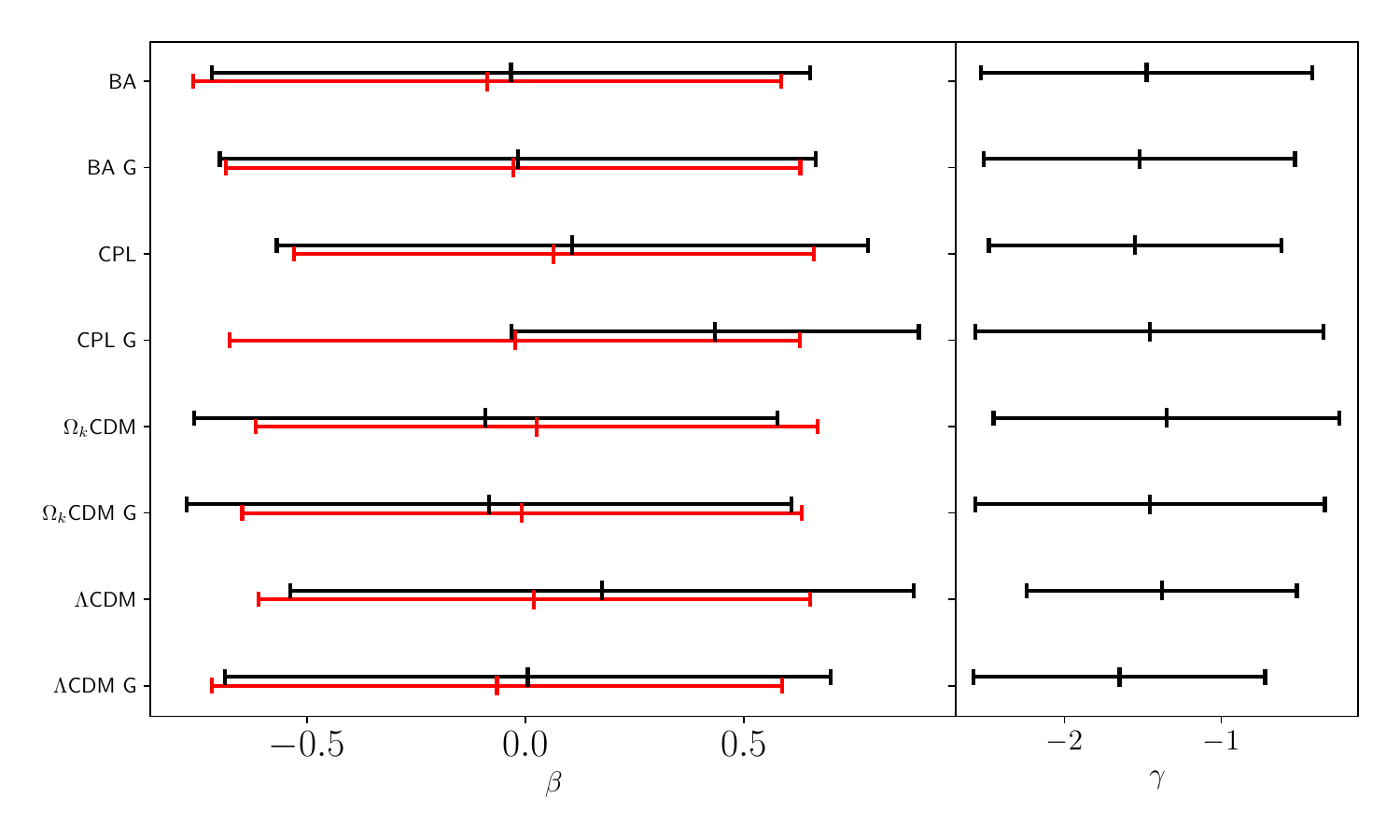}
    \includegraphics[width=0.47\textwidth]{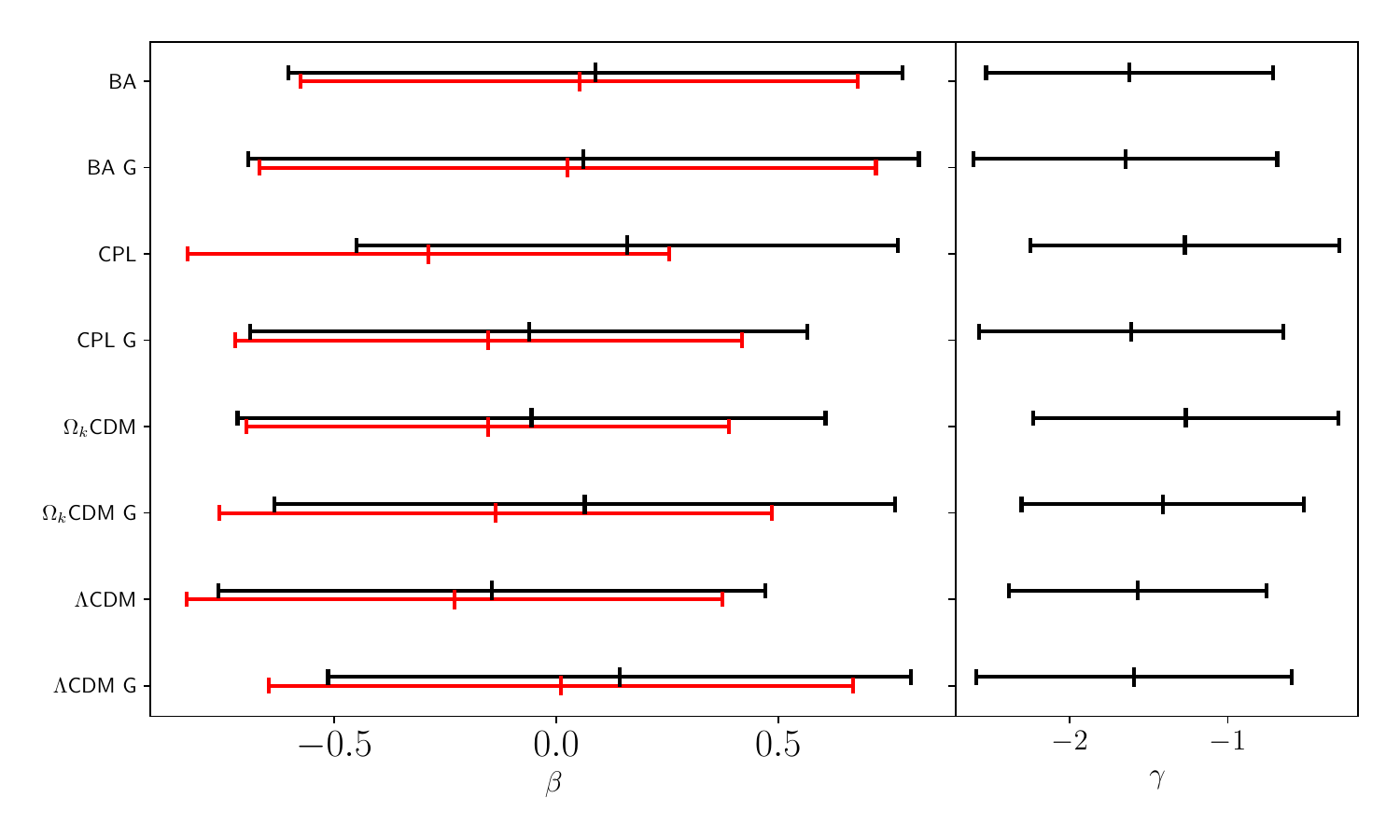}
    \caption{Mean values and errors (68\%CL) for the parameters $\beta$ and $\gamma$ for the standard approximation (red) and the extended one (black) different models. To~the (\textbf{left}) are the results for TD1 and to the (\textbf{right})---from~TD2.}
    \label{fig:Vard_Xiao_beta_gamma}
\end{figure}

Finally, we are going to discuss the LIV parameters, $\alpha, \beta$ and $\gamma$. They are shown in Figures~\ref{fig:Vard_Xiao_alpha} and \ref{fig:Vard_Xiao_beta_gamma}. In~Figure~\ref{fig:Vard_Xiao_alpha},  we can see the values for $\alpha$ we obtained (where it should be noted that the lower bound on $\alpha$ is, of~course, $0$). As~a whole, TD1 gives 10 times higher values for $\alpha$ than $TD2$ (which has been noticed also in our previous paper). We also see that the extended models do not have a conclusive improvement over the standard approximation, even though, for~some of the models considered, they lead to significantly smaller $\alpha$, meaning higher $E_{QG}$.

In Figure~\ref{fig:Vard_Xiao_beta_gamma}, we show the parameters $\beta$ and $\gamma$. We note that they are mostly unconstrained in all of the models and while there are differences between the standard and the extended approximations, they are minor. {{The} posteriors for the LIV parameters can be found in {Appendix} \ref{apendix:2}.
}
\vspace{-6pt}

To obtain the values for $E_{QG}$, {{one} needs to use the formula}  $E_{QG}=\Delta E/(H_0 a_{LIV})$.  {This requires inputting the specific energy band for each dataset ($\Delta E$), and~choosing a value for $H_0$. The~error in this case will be $$\sigma(E_{EQ})=E_{QG} \sqrt{\frac{{\sigma_{H0}}^{2}}{{H_0}^{2}}+\frac{{\sigma_{\alpha}}^{2}}{\alpha^{2}}}.$$}
 
 {{Since} $\alpha=0$ is the low limit of our priors, which gives an upper bound for $E_{QG}$ infinity,  we can get only the lower bound of the quantum gravity energy. We take as $\sigma_{\alpha}$ the 68\% CL corresponding to $1\sigma$ deviation and we take the minimal and the maximal energies obtained for the different models. The~results can be found in Table~\ref{tab:Eqg}. In~it, one needs to remember that the error comes  from the incertitude in $H_0$ and the numerical estimation of $\alpha$; thus, it is not supposed to be taken as a measurement, but~as an estimation of the error we get from this study.  }
\begin{table}[H]
\tablesize{\footnotesize}
     \caption{{The minimal}  and maximal energy $E_{QG}$ for the standard approximation (SA) and the extended approximation (EA).}
    \label{tab:Eqg}
  \newcolumntype{C}{>{\centering\arraybackslash}X}
\setlength{\tabcolsep}{2mm}
\begin{tabularx}{\textwidth}{ccccc}
\toprule
        \textbf{Dataset} & \textbf{{\boldmath$E_{QG}^{min, SA}\times 10^{17}$Gev}}& \textbf{{\boldmath$E_{QG}^{max, SA}\times 10^{17}$Gev}} & \textbf{{\boldmath$E_{QG}^{min, EA}\times 10^{17}$Gev}} & \textbf{{\boldmath$E_{QG}^{max, EA}\times 10^{17}$Gev}} \\
        \midrule
        $H_0 = 73.04 \pm 1.04$ & & & & \\
        \midrule
        TD1 &  $1.14 \pm  0.84$& $0.81 \pm  0.57$&  $1.39 \pm  1.01$&  $0.93 \pm  0.68$\\
        TD2 & $48.0 \pm  35.6$& $35.5 \pm 25.5$& $74.6 \pm 55.9$& $35.8 \pm  27.1$\\
        \midrule
        $H_0 = 67.4\pm 0.5$ & & & & \\
        \midrule
        TD1 &$1.24\pm0.91$ & $0.88\pm 0.62$ & $ 1.51\pm1.09$ & $1.01\pm0.735$ \\
        TD2 & $ 48.0\pm35.6$ & $ 35.5\pm25.5$ & $ 74.6\pm55.9$ & $ 35.8\pm27.1$\\ 
        \bottomrule
    \end{tabularx}
   
\end{table}
\unskip

\section{Discussion}
\label{sec:discussion}
We have studied the LIV bounds and the effect of  cosmology based on two different datasets TD1~\cite{Bernardini:2014oya,Vardanyan:2022ujc} and TD2~\cite{Xiao:2022ovb}. To~allow testing for new approximations for the intrinsic time delay, we have added two extended models---the intrinsic GRB lag-luminosity relation approximation for TD1 and the energy-dependent approximation for TD2 and we have compared them to the standard approximation (``constant'' intrinsic time, i.e.,~not depending on the energy or the luminosity). To~study the effect of the cosmological model, we have used additional cosmological datasets including transversal BAO, the~Pantheon Plus dataset and the CMB distance priors. These are some of the most robust cosmological datasets; thus, they provide the best opportunity to study the joint effect of cosmology and LIV.  We have considered $\Lambda$CDN, $\Omega_K$CDM, the~CPL and the BA dark energy~models.

From the results we obtain, we see that the strongest effect is due to the prior on the parameter $\frac{c}{H_0 r_d}$ and not so much on the approximation for the intrinsic lag. This is because first the LIV effect is expected to be very small, and~second, the~intrinsic lag parameters are largely unbound from the current data. Instead, we see that the results for $\alpha$ depend a lot on the cosmological model. The~lowest bounds of $E_{QG}$ are for $\Omega_K$CDM and the highest for $\Lambda$CDM G (TD1) and BA (TD2). Surprisingly, despite the new degrees of freedom introduced by the LIV parameters, the~$\Omega_K$CDM model suggests a closed universe. In~terms of LIV energy, we obtain as lowest bound for TD1 $E_{QG}>0.8 \times 10^{17}$ GeV and for TD2: $E_{QG}>3.5 \times 10^{18}$ GeV. In~both cases, the~error is significant, regardless of the small error of $H_0$. TD1 tends to give 10 times larger values of $\alpha$ than TD2. The~effect of the TD datasets on the cosmological parameters becomes noticeable if $\alpha$ is an order higher than the inferred one, meaning lower than the currently estimated $E_{QG}$.

In conclusion, we see that for the moment, the~time delay datasets are not precise enough to constrain the cosmological effects, while the cosmological models have a serious effect on the LIV constraints. For~this situation to change, we need to improve the model of the GRB central engine and also to be able to better constrain the propagational effects, which are considered negligible at high energies; however, one needs to remember that not all measurements are made at very high energies, especially for the older GRB collections. Finally, a~new and better approximation for the intrinsic time delay could benefit both GRB theoretical models and cosmological~studies.

\vspace{6pt} 

%

\funding{{This work was conducted in a Short Term Scientific Missions (STSM) in Spain, funded by the COST Action CA21136 ``Addressing observational tensions in cosmology with systematics and fundamental physics (CosmoVerse)''. }
}

\acknowledgments{D.S. is grateful to Diego Rubiera-Garcia for his hospitality at Complutense~University. }

\dataavailability{Data are contained within the article.}

\conflictsofinterest{{The author declares no conflicts of interest.}
} 

\appendixtitles{yes} 
\appendixstart
\appendix
\section[\appendixname~\thesection]{Posteriors of the~Models}
\label{apendix:1}


\begin{figure}[H]
    \includegraphics[width=0.43\textwidth]{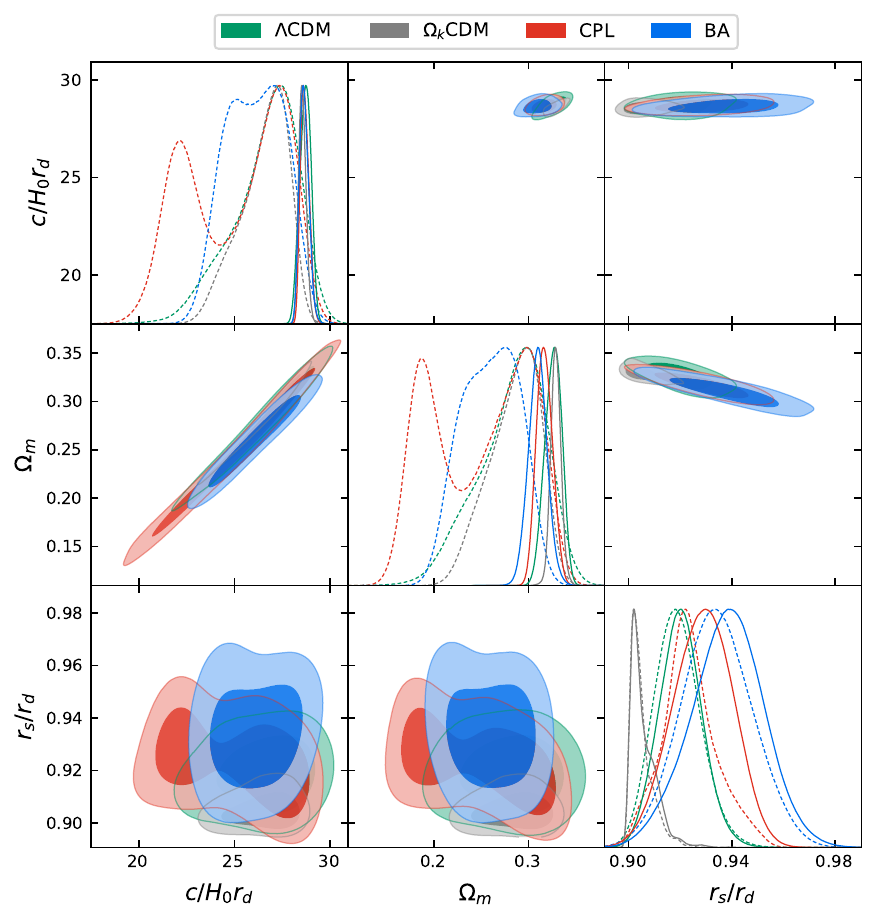}
    \includegraphics[width=0.43\textwidth]{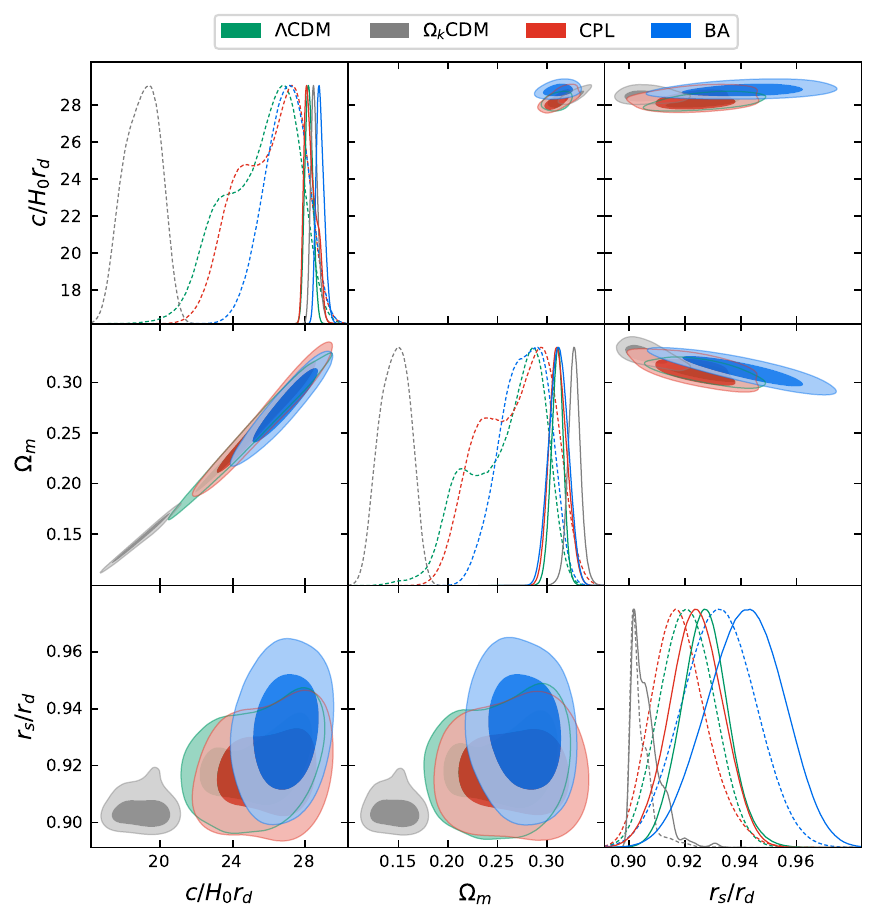}
    \caption{Posteriors for the TD1 dataset for the standard approximation (\textbf{left panel}) and the extended approximation (\textbf{right panel}). On~each panel, the~uniform prior is on the bottom triangle plot, the~Gaussian---on the top~one. }
    \label{fig:TD1_all}
\end{figure}
\unskip

\begin{figure}[H]
    \includegraphics[width=0.43\textwidth]{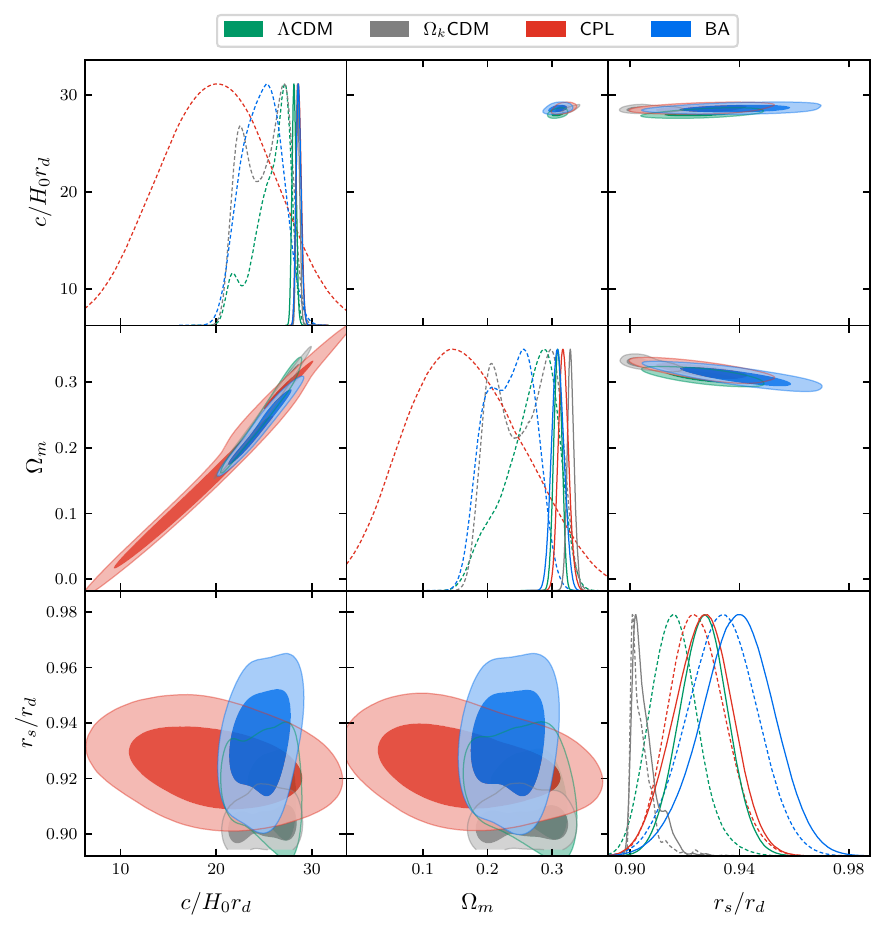}
    \includegraphics[width=0.43\textwidth]{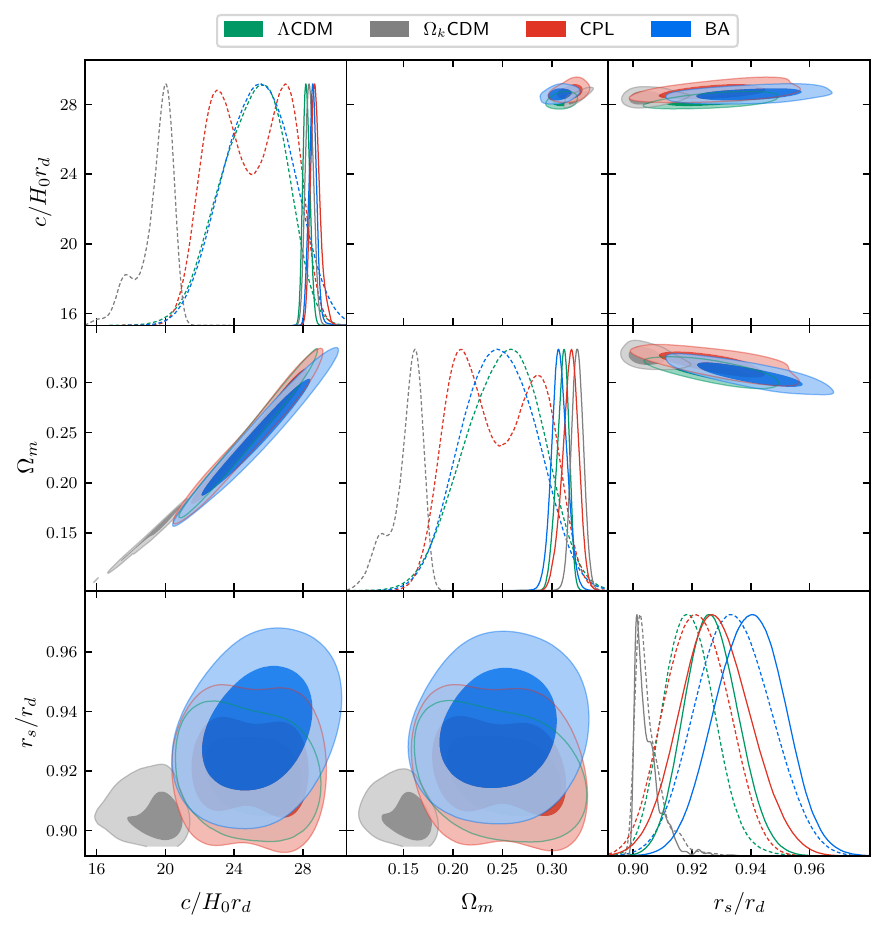}
    \caption{Posteriors for the TD2 dataset for the standard approximation (\textbf{left panel}) and the extended approximation (\textbf{right panel}). On~each panel, the~uniform prior is on the bottom triangle plot, the~Gaussian---on the top~one. }
    \label{fig:TD2_all}
\end{figure}

\section[\appendixname~\thesection]{LIV Parameters~Posteriors}
\label{apendix:2}

\begin{figure}[H]
    \includegraphics[width=0.3\textwidth]{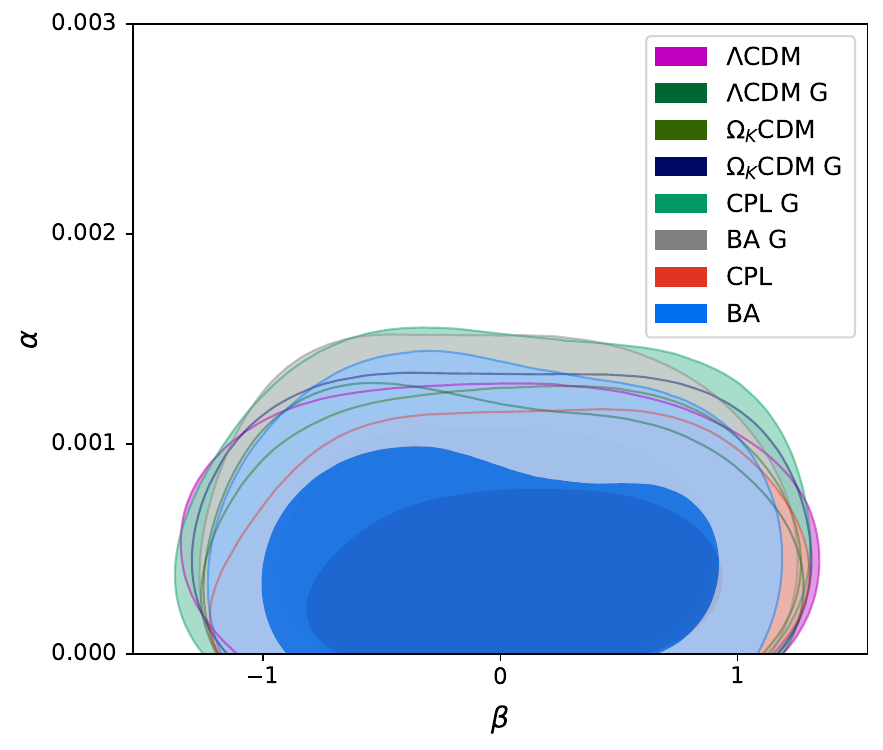}
    \includegraphics[width=0.69\textwidth]{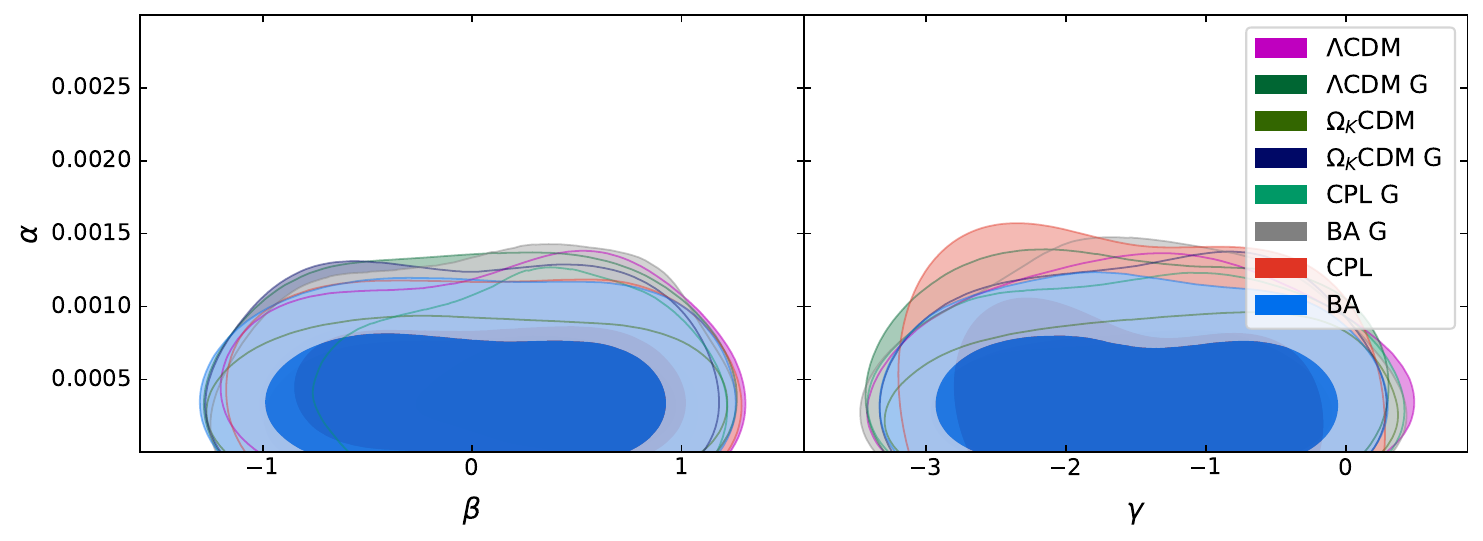}\\
    \includegraphics[width=0.3\textwidth]{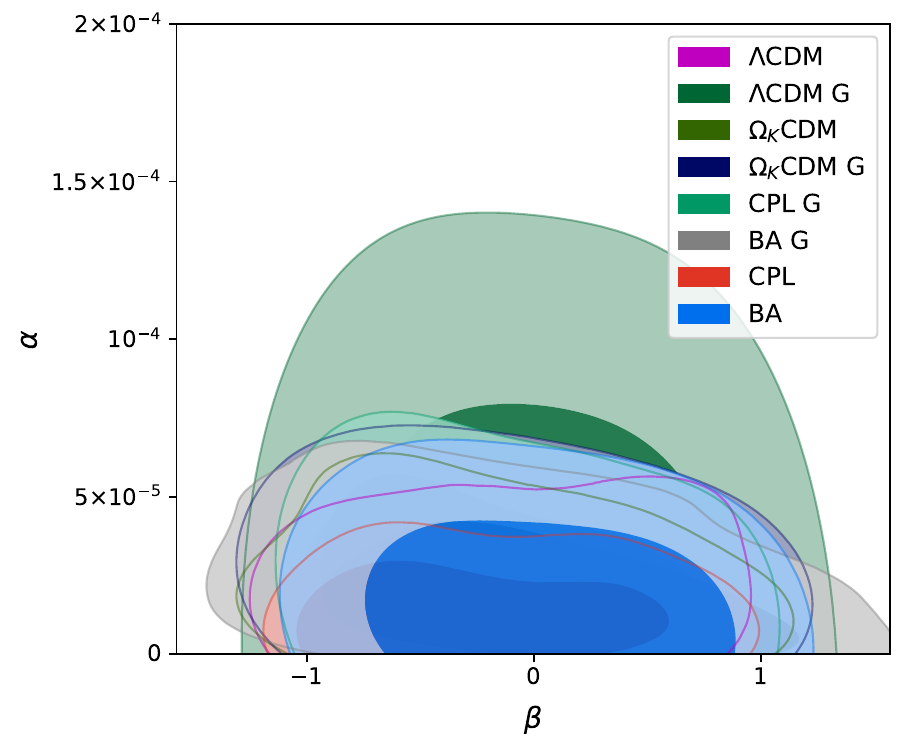}
    \includegraphics[width=0.69\textwidth]{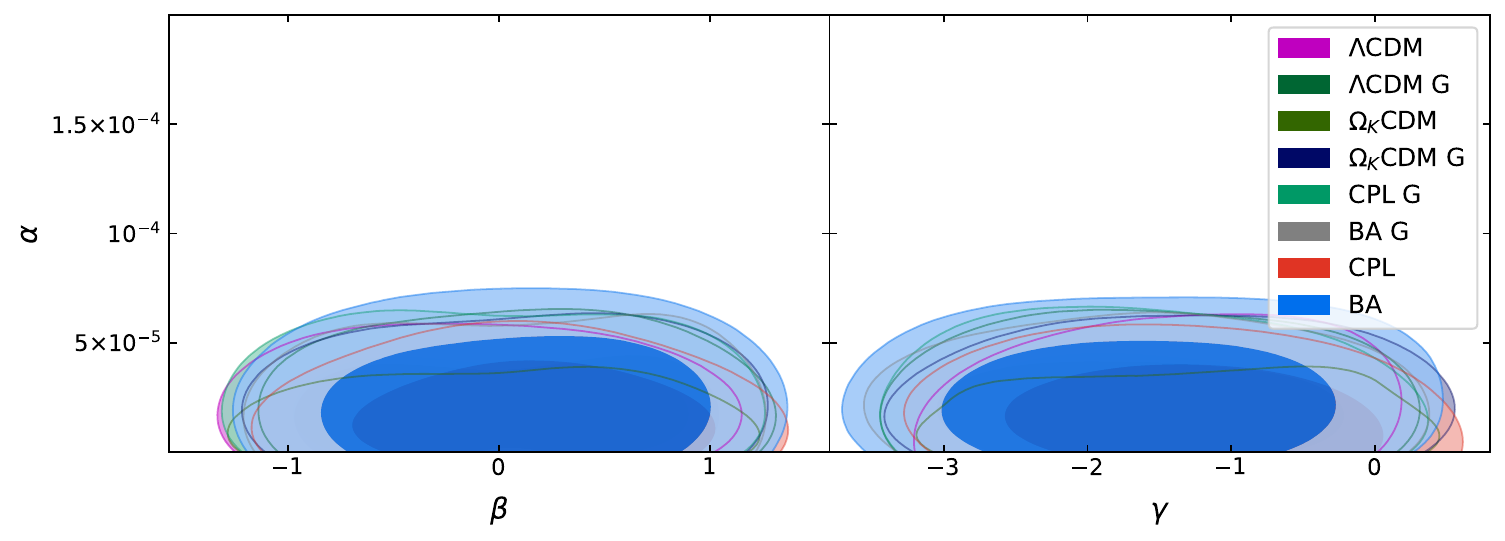}
    \caption{{{Posteriors} (95\% CL and 68\% CL) for the TD1 dataset (\textbf{top panel}) and the TD2 dataset (\textbf{bottom panel}). On~the left, we have the SA case with its two parameters $\alpha$ and $\beta$, on~the right, the~EA case with its three parameters $\alpha$, $\beta$ and $\gamma$. As~seen for the plots, only $\alpha$ can be bounded with these datasets.} }
    \label{fig:TD_liv}

\end{figure}

\section[\appendixname~\thesection]{Tables of the~Results}
\label{apendix:3}

\begin{table}[H]
\tablesize{\footnotesize}
	\caption{{Mean}  values for the TD1 dataset for the standard~approximation.}
\setlength{\tabcolsep}{2.8mm}
		\newcolumntype{C}{>{\centering\arraybackslash}X}
\begin{tabularx}{\textwidth}{ccccccc}
			\toprule
			\textbf{Model} & \boldmath{$c/H_0 r_d$} & \boldmath{$\Omega_K$} &\boldmath{ $\Omega_m$} & \textbf{$r_s/r_d$} & \textbf{$w_0$} & \textbf{$w_a$} \\
			\midrule
			\midrule
			$\Lambda$CDM G & $28.7\pm 0.3$ & 0.000 & $0.33\pm 0.01$ & $0.92\pm 0.01$ & 0.000 & 0.000 \\
			\midrule
			$\Lambda$CDM & $26.7\pm 1.9$ & 0.000 & $0.28\pm 0.04$ & $0.92\pm 0.01$ & 0.000 & 0.000 \\
			\midrule
			$\Omega_K$CDM G & $28.6\pm 0.2$ & $-9.95\pm 1.81$ & $0.33\pm 0.01$ & $0.91\pm 0.0$ & 0.000 & 0.000 \\
			\midrule
			$\Omega_K$CDM & $26.6\pm 1.4$ & $-7.30\pm 1.85$ & $0.29\pm 0.03$ & $0.9\pm 0.0$ & 0.000 & 0.000 \\
			\midrule
			CPL G & $28.7\pm 0.2$ & 0.000 & $0.32\pm 0.01$ & $0.93\pm 0.01$ & $-0.84\pm 0.03$ & $-0.3\pm 0.15$ \\
			\midrule
			CPL & $25.1\pm 3.2$ & 0.000 & $0.25\pm 0.06$ & $0.92\pm 0.01$ & $-0.84\pm 0.03$ & $-0.33\pm 0.13$ \\
			\midrule
			BA G & $28.6\pm 0.2$ & 0.000 & $0.31\pm 0.01$ & $0.94\pm 0.01$ & $-0.8\pm 0.05$ & $-0.31\pm 0.14$ \\
			\midrule
			BA & $26.1\pm 1.7$ & 0.000 & $0.26\pm 0.04$ & $0.93\pm 0.01$ & $-0.8\pm 0.04$ & $-0.34\pm 0.13$ \\
			\bottomrule
		\end{tabularx}
\end{table}
\unskip

\begin{table}[H]
\tablesize{\footnotesize}
\caption{{Mean} values for the TD1 dataset for the extended~approximation.}
\setlength{\tabcolsep}{2.8mm}
		\newcolumntype{C}{>{\centering\arraybackslash}X}
\begin{tabularx}{\textwidth}{ccccccc}
			\toprule
			\textbf{Model} & \boldmath{$c/H_0 r_d$} & \boldmath{$\Omega_K$} &\boldmath{ $\Omega_m$} & \textbf{$r_s/r_d$} & \textbf{$w_0$} & \textbf{$w_a$} \\
			\midrule
			$\Lambda$CDM GL & $28.2\pm 0.2$ & 0.000 & $0.31\pm 0.01$ & $0.93\pm 0.01$ & 0.000 & 0.000 \\
			\midrule
			$\Lambda$CDM L & $25.5\pm 2.5$ & 0.000 & $0.26\pm 0.05$ & $0.92\pm 0.01$ & 0.000 & 0.000 \\
			\midrule
			$\Omega_K$CDM GL & $28.5\pm 0.2$ & $-9.44\pm 1.97$ & $0.33\pm 0.01$ & $0.91\pm 0.0$ & 0.000 & 0.000 \\
			\midrule
			$\Omega_K$CDM L & $19.0\pm 1.1$ & $-3.84\pm 9.06$ & $0.15\pm 0.02$ & $0.9\pm 0.0$ & 0.000 & 0.000 \\
			\midrule
			CPL GL & $28.3\pm 0.3$ & 0.000 & $0.31\pm 0.01$ & $0.92\pm 0.01$ & $-0.84\pm 0.03$ & $-0.32\pm 0.14$ \\
			\midrule
			CPL L & $26.0\pm 2.1$ & 0.000 & $0.27\pm 0.04$ & $0.92\pm 0.01$ & $-0.84\pm 0.03$ & $-0.32\pm 0.13$ \\
			\midrule
			BA GL & $28.8\pm 0.2$ & 0.000 & $0.31\pm 0.01$ & $0.94\pm 0.01$ & $-0.79\pm 0.04$ & $-0.33\pm 0.12$ \\
			\midrule
			BAL & $26.9\pm 1.2$ & 0.000 & $0.28\pm 0.02$ & $0.93\pm 0.01$ & $-0.81\pm 0.05$ & $-0.3\pm 0.14$ \\
			\bottomrule
		\end{tabularx}
	
\end{table}
\unskip

\begin{table}[H]
\tablesize{\footnotesize}
\caption{{Mean} values for the TD2 dataset for the standard~.}
\setlength{\tabcolsep}{2.8mm}
		\newcolumntype{C}{>{\centering\arraybackslash}X}
\begin{tabularx}{\textwidth}{ccccccc}
			\toprule
			\textbf{Model} & \boldmath{$c/H_0 r_d$} & \boldmath{$\Omega_K$} &\boldmath{ $\Omega_m$} &\textbf{$r_s/r_d$} & \textbf{$w_0$} & \textbf{$w_a$} \\
			\midrule
			$\Lambda$CDM G & $28.1\pm 0.2$ & 0.000 & $0.31\pm 0.01$ & $0.93\pm 0.01$ & 0.000 & 0.000 \\
			\midrule
			$\Lambda$CDM & $25.7\pm 1.9$ & 0.000 & $0.27\pm 0.04$ & $0.92\pm 0.01$ & 0.000 & 0.000 \\
			\midrule
			$\Omega_K$CDM G & $28.5\pm 0.2$ & $-9.85\pm 1.82$ & $0.33\pm 0.01$ & $0.91\pm 0.0$ & 0.000 & 0.000 \\
			\midrule
			$\Omega_K$CDM & $25.0\pm 2.7$ & $-6.73\pm 2.27$ & $0.25\pm 0.05$ & $0.9\pm 0.0$ & 0.000 & 0.000 \\
			\midrule
			CPL G & $28.6\pm 0.3$ & 0.000 & $0.32\pm 0.01$ & $0.93\pm 0.01$ & $-0.83\pm 0.03$ & $-0.34\pm 0.13$ \\
			\midrule
			CPL & $19.9\pm 6.9$ & 0.000 & $0.17\pm 0.1$ & $0.92\pm 0.01$ & $-0.84\pm 0.03$ & $-0.35\pm 0.11$ \\
			\midrule
			BA G & $28.6\pm 0.2$ & 0.000 & $0.31\pm 0.01$ & $0.94\pm 0.01$ & $-0.8\pm 0.05$ & $-0.31\pm 0.13$ \\
			\midrule
			BA & $24.7\pm 2.3$ & 0.000 & $0.23\pm 0.04$ & $0.93\pm 0.01$ & $-0.81\pm 0.05$ & $-0.31\pm 0.14$ \\
			\bottomrule
		\end{tabularx}
	
\end{table}
\unskip

\begin{table}[H]
\tablesize{\footnotesize}
	\caption{{Mean} values for the TD2 dataset for the extended~approximation.}
\setlength{\tabcolsep}{2.8mm}
		\newcolumntype{C}{>{\centering\arraybackslash}X}
\begin{tabularx}{\textwidth}{ccccccc}
			\toprule
		\textbf{Model} & \boldmath{$c/H_0 r_d$} & \boldmath{$\Omega_K$} &\boldmath{ $\Omega_m$} & \textbf{$r_s/r_d$} & \textbf{$w_0$} & \textbf{$w_a$} \\
			\midrule
			$\Lambda$CDM GL & $28.2\pm 0.2$ & 0.000 & $0.31\pm 0.01$ & $0.93\pm 0.01$ & 0.000 & 0.000 \\
			\midrule
			$\Lambda$CDM L & $25.1\pm 2.1$ & 0.000 & $0.25\pm 0.05$ & $0.92\pm 0.01$ & 0.000 & 0.000 \\
			\midrule
			$\Omega_K$CDM GL & $28.4\pm 0.2$ & $-9.26\pm 1.63$ & $0.33\pm 0.01$ & $0.9\pm 0.0$ & 0.000 & 0.000 \\
			\midrule
			$\Omega_K$CDM L & $19.5\pm 1.2$ & $-4.04\pm 1.09$ & $0.15\pm 0.02$ & $0.91\pm 0.0$ & 0.000 & 0.000 \\
			\midrule
			CPL GL & $28.7\pm 0.3$ & 0.000 & $0.32\pm 0.01$ & $0.93\pm 0.01$ & $-0.82\pm 0.03$ & $-0.33\pm 0.13$ \\
			\midrule
			CPL L & $24.9\pm 2.5$ & 0.000 & $0.25\pm 0.05$ & $0.92\pm 0.01$ & $-0.83\pm 0.03$ & $-0.34\pm 0.12$ \\
			\midrule
			BA GL & $28.6\pm 0.2$ & 0.000 & $0.31\pm 0.01$ & $0.94\pm 0.01$ & $-0.79\pm 0.04$ & $-0.34\pm 0.11$ \\
			\midrule
			BAL & $25.4\pm 2.1$ & 0.000 & $0.25\pm 0.04$ & $0.93\pm 0.01$ & $-0.8\pm 0.05$ & $-0.32\pm 0.13$ \\
			\bottomrule
		\end{tabularx}
\end{table}
\unskip

\begin{table}[H]
  \caption{{The}  LIV parameters for TD1: to the left---the standard approximation, to~the right---the extended~one. }
 \newcolumntype{C}{>{\centering\arraybackslash}X}
\setlength{\tabcolsep}{4mm}
\begin{tabularx}{\textwidth}{cccccc}
			\toprule
			\textbf{Model} & \textbf{B} & \boldmath{$\beta$} &\boldmath{$\alpha \times 10^{-4}$} & \boldmath{$\beta$} & \boldmath{$\gamma$}\\
			\midrule
			$\Lambda$CDM G & $4.01\pm 2.82$ & $-0.06\pm 0.65$ & $4.56\pm 3.34$ & $0.006\pm 0.7$ & $-1.65\pm 0.93$\\
			\midrule
			$\Lambda$CDM & $4.46\pm 3.08$ & $0.02\pm 0.63$&$4.34\pm 2.87$ & $0.18\pm 0.71$ & $-1.38\pm 0.86$ \\
			\midrule
			$\Omega_K$CDM G & $4.73\pm 3.36$ & $-0.008\pm 0.64$&$4.27\pm 3.18$ & $-0.08\pm 0.69$ & $-1.45\pm 1.11$ \\
			\midrule
			$\Omega_K$CDM & $4.11\pm 3.02$ & $0.03\pm 0.64$& $3.04\pm 2.21$ & $-0.09\pm 0.67$ & $-1.35\pm 1.1$  \\
			\midrule
			CPL G & $5.21\pm 3.66$ & $-0.02\pm 0.65$ & $3.85\pm 2.72$ & $0.43\pm 0.47$ & $-1.46\pm 1.11$\\
			\midrule
			CPL & $3.69\pm 2.69$ & $0.07\pm 0.6$& $4.26\pm 3.06$ & $0.11\pm 0.68$ & $-1.55\pm 0.93$  \\
			\midrule
			BA G & $5.06\pm 3.95$ & $-0.03\pm 0.66$ & $3.99\pm 3.10$ & $-0.02\pm 0.68$ & $-1.52\pm 0.99$\\
			\midrule
			BA & $4.57\pm 3.32$ & $-0.09\pm 0.67$ &$3.86\pm 3.00$ & $-0.03\pm 0.69$ & $-1.48\pm 1.06$\\
			\bottomrule
		\end{tabularx}
		  \end{table}
\unskip
		

\begin{table}[H]
\caption{{The}  LIV parameters for TD2: to the left---the standard approximation, to~the right---the extended~one.}
\newcolumntype{C}{>{\centering\arraybackslash}X}
\setlength{\tabcolsep}{4mm}
\begin{tabularx}{\textwidth}{cccccc}
			\toprule
			\textbf{Model} & \boldmath{$\alpha  \times 10^{-5}$} & \boldmath{$\beta$}&\boldmath{$\alpha \times 10^{-5}$} & \boldmath{$\beta$} & \boldmath{$\gamma$}  \\
			\midrule
			$\Lambda$CDM G & $2.03\pm 1.49$ & $0.01\pm 0.66$& $2.15\pm 1.64$ & $0.14\pm 0.66$ & $-1.59\pm 1.0$\\
			\midrule
			$\Lambda$CDM & $1.75\pm 1.25$ & $-0.23\pm 0.6$& $1.90\pm 1.45$ & $-0.14\pm 0.62$ & $-1.57\pm 0.81$ \\
			\midrule
			$\Omega_K$CDM G & $2.35\pm 1.69$ & $-0.14\pm 0.62$ & $2.15\pm 1.48$ & $0.06\pm 0.7$ & $-1.41\pm 0.89$ \\
			\midrule
			$\Omega_K$CDM & $1.79\pm 1.07$ & $-0.15\pm 0.54$ &$1.12\pm 0.84$ & $-0.06\pm 0.66$ & $-1.27\pm 0.96$\\
			\midrule
			CPL G & $2.22\pm 1.60$ & $-0.15\pm 0.57$ &$2.10\pm 1.57$ & $-0.06\pm 0.63$ & $-1.61\pm 0.96$\\
			\midrule
			CPL & $1.23\pm 0.96$ & $-0.29\pm 0.54$& $1.71\pm 1.27$ & $0.16\pm 0.61$ & $-1.27\pm 0.98$\\
			\midrule
			BA G & $1.97\pm 1.47$ & $0.03\pm 0.69$ & $2.00\pm 1.39$ & $0.06\pm 0.76$ & $-1.65\pm 0.96$ \\
			\midrule
			BA & $1.74\pm 1.29$ & $0.05\pm 0.63$ &$2.33\pm 1.76$ & $0.09\pm 0.69$ & $-1.62\pm 0.91$ \\
			\bottomrule
		\end{tabularx}
  
\end{table}

\begin{adjustwidth}{-\extralength}{0cm}
\setenotez{list-name=Note}
\printendnotes[custom]

\reftitle{References}

\PublishersNote{}
\end{adjustwidth}

\end{document}